\documentclass[
aip,
superscriptaddress, 
reprint, 
onecolumn
]{revtex4-1} 
\usepackage{natbib}
\usepackage{graphicx}
\usepackage{wrapfig} 
\usepackage{color}
\usepackage{hyperref}
\usepackage{amsmath, amssymb}

\begin{document}

\title{
Linear control theory for jammed particle systems
}

\author{Erin G. Teich}
\affiliation{Department of Physics and Astronomy, Wellesley College, Wellesley, MA 02481 USA}
\affiliation{To whom correspondence should be addressed: et106@wellesley.edu}

\author{Jason Z. Kim}
\affiliation{Department of Physics, Cornell University, Ithaca, NY 14853 USA}

\author{Dani S. Bassett}
\affiliation{Department of Bioengineering, School of Engineering \& Applied Science, University of Pennsylvania, Philadelphia, PA 19104 USA}
\affiliation{Department of Physics \& Astronomy, College of Arts \& Sciences, University of Pennsylvania, Philadelphia, PA 19104 USA}
\affiliation{Department of Electrical \& Systems Engineering, School of Engineering \& Applied Science, University of Pennsylvania, Philadelphia, PA 19104 USA}



\begin{abstract} 
Amorphous particulate matter constitutes a wide range of natural and synthetic materials. 
Despite this ubiquity, the way in which these systems’ disordered microstructure couples to their often subtle and complex dynamical behavior is not yet fully understood, with profound consequences for phenomena ranging from landscape evolution to cellular unjamming during tumor metastasis. 
With this paper, we introduce tools from linear control theory that quantify system response to external input, and demonstrate their utility in elucidating the dynamics of jammed amorphous materials under stress. 
Our results indicate that average controllability, the response of a system to perturbation, strongly correlates with particle rearrangement in systems subject to quasistatic shear, implying that average controllability is an accurate predictor of rearrangement dynamics in certain contexts.
Moreover, we show that the time scale over which average controllability is calculated can be tuned to optimize its predictive capacity for particle rearrangement.
Values of the optimal time scale provide physical insight into the system; namely, that multiple rearranging particles participate on average in vibrational eigenmodes of lower and lower energy as the system is sheared until the rearrangement event.
Broadly, our study demonstrates that linear control theory is a promising mathematical framework for predicting and designing mechanical response in disordered media.
\end{abstract}

\date{\today}

\maketitle

\section{Introduction}
Disordered solids are ubiquitous in nature and our everyday lives.
This diverse set of materials encompasses systems ranging from the ground under our feet, to the foam on the surface of our espresso drinks, to the metallic glasses under active development for their unique mechanical properties \cite{Katgert2013,Nicolas2018}.
Like all solids, these materials elastically deform under applied loading until a threshold stress is reached, at which point they macroscopically yield \cite{Nicolas2018}.
The successful prediction of where and how this yielding occurs is crucial for the design and manufacture of failure-resistant materials, and can even help in understanding landscape evolution \cite{Ferdowsi2017,Jerolmack2019} and the way cancer spreads in the body \cite{Oswald2017,Gottheil2023}.

Where in disordered materials is yielding likely to begin?
It has long been hypothesized \cite{Spaepen1977,Argon1979,Falk1998} that localized particle rearrangement occurs in disordered solids under sufficient external loading prior to macroscopic yielding, and these rearrangements have been observed in Lennard-Jones glasses \cite{Maloney2006,Tanguy2006}, metallic glasses \cite{Shang2014,Dong2023}, jammed packings of soft repulsive particles \cite{Manning2011}, dense colloidal suspensions \cite{Schall2007,Chen2010c,Jensen2014,Keim2014}, emulsions \cite{Desmond2015}, foams \cite{Debregeas2001,Biance2009}, and granular materials \cite{Hayman2011,Amon2013,Keim2013,Keim2014,Cai2021,Kozlowski2021}.
Unfortunately, the generalized prediction of where these rearrangements occur in each type of system is far from straightforward, due to these materials' disordered microstructure and the consequently complicated relationship between the dynamics of yield and material structure.

Many methods have been developed for the prediction of rearrangements in disordered systems under perturbation.
These methods include those based solely on structural features such as bond-orientational order \cite{Kawasaki2007,Keim2014}, energetically favored motifs \cite{Malins2013a}, and steric hindrance \cite{Tong2018}, those based on the vibrational modes of the system under a linear approximation \cite{Widmer-Cooper2008,Tanguy2010,Manning2011,Chen2011,Tong2014}, those based on the nonlinear response of the system \cite{Gartner2016,Kapteijns2020}, those utilizing machine-learned system features \cite{Cubuk2015,Schoenholz2017,Boattini2020,Bapst2020}, and those directly measuring residual plastic strength under local shearing \cite{Patinet2016}.
Recently, many of these metrics were tested in a comparative study \cite{Richard2020a} that evaluated their capacity to predict particle rearrangement in two-dimensional model glasses.
This study found that the metrics had a range of predictive capabilities across different glass models and for different types of rearrangements, with metrics based on the linear response of the material generally performing quite well in predicting the first rearrangement in brittle and ductile systems and subsequent rearrangements in ductile systems.
The relative success of the metrics based on linear response indicates that a broad theoretical framework grounded in the linearized dynamics of amorphous solids will be useful in many contexts; however, to our knowledge, no such general framework has ever been introduced for this problem.

Here, we propose that control theory \cite{Schulz2006} (and specifically linear control theory \cite{Kailath1980}) is a general theoretical framework useful for the prediction and eventual design of failure in amorphous solids.
Control theory is a mature analytical toolkit employed extensively to understand and control response in complex dynamical systems in diverse contexts ranging from robotics \cite{Wit2012}, to complex biochemical networks \cite{Iglesias2010}, to marvels of engineering like the interferometer used to detect gravitational waves in the LIGO experiment \cite{Bechhoefer2005}, to network models of the brain \cite{Parkes2024}.
In this paper, we investigate one metric derived from control theory---\emph{average controllability} \cite{Summers2014,Pasqualetti2014}, or the trace of the controllability Gramian---and we demonstrate its use in understanding material failure.
Average controllability quantifies the response of the system over some time horizon to impulse input to one or more of its degrees of freedom.
This metric has been employed in a wide variety of control problems, including the choice of optimal leaders in multi-agent networks\cite{Fitch2016}, optimal actuators for the injection of current into the European power grid\cite{Summers2014a}, optimal green light scheduling at intersections to improve traffic congestion\cite{Bianchin2018}, optimal communication networking to guard flocks of UAVs against wind gusts\cite{Chapman2011}, and optimal distributions of piezoelectric material to control vibrations in a cantilever beam\cite{Guzman2020}.
Average controllability has also been used in contexts far beyond traditional control, including studies of optimal intervention strategies during drinking events\cite{GonzalezVillasanti2019}
and studies of network dynamics in the brain \cite{Gu2015}. 

In this work, we show that average controllability is a successful and informative predictor of rearrangements in amorphous solids under quasistatic shear.
We simulate jammed systems, in which rigidity is induced via applied stress \cite{Cates1998, Liu2010b}, and find that particle propensity to rearrange under athermal, quasistatic shear \cite{Malandro1999,Maloney2006} is strongly correlated with the average controllability of the system when that particle receives impulse input.
This correlation is high for multiple definitions of rearrangement, indicating that average controllability is an effective predictor of particle rearrangement in jammed systems under quasistatic shear.
In particular, the predictive capability of average controllability when it is calculated over long time horizons is on par with that of another metric based on linear response, vibrality \cite{Tong2014}, which was shown to be a successful rearrangement predictor with respect to several other metrics in various glass-former models \cite{Richard2020a}.
Moreover, we show that the predictive capacity of average controllability can be enhanced by tuning the time horizon over which average controllability is calculated.
This tuning process provides physical insight into the system on approach to rearrangement.
In particular, average values of the optimal time horizon indicate that multiple rearranging particles participate in vibrational eigenmodes of lower and lower energy as the system approaches the rearrangement event.
Broadly, our paper demonstrates the potential benefits of using control theory to better understand disordered systems under external loading. 
Further, it represents a first step toward the ultimate goal of the design of failure in disordered materials using a control theoretic framework.

\section{Methods}
In order to investigate the efficacy of tools from control theory to predict rearrangement in jammed disordered solids, we prepare an ensemble of jammed disordered solids, shear each solid to induce rearrangements, and quantify which particles participate in those rearrangements.
We then calculate average controllability as well as vibrality for all particles in the system, and comparatively evaluate the predictive capabilities of these metrics for describing particle rearrangement.

\subsection{Preparing jammed systems and inducing rearrangements}

All systems are two-dimensional bidisperse mixtures of 2500 particles consisting of equal numbers of small and large particles, where the ratio of particle diameters is 1:1.4, chosen to suppress crystallization.
Particles interact via a Hertzian repulsion:
\begin{align*}
    U_{ij}(r) &= 
    \begin{cases}
    \epsilon \left( 1 - \frac{r}{\sigma_{ij}} \right)^{\frac{5}{2}} & \text{if } r < \sigma_{ij}\\
    0 & \text{if } r \geq \sigma_{ij}~.
    \end{cases}
\end{align*}
In the above, $U_{ij}(r)$ is the potential energy corresponding to two particles of species $i$ and $j$, respectively, separated by center-to-center distance $r$.
The factor $\epsilon = 1$ sets the energy scale.
Species $i, j \in \{A,B\}$, where $\sigma_A = 1$ sets the length scale in the system, and $\sigma_B = 1.4$.
The composite diameter $\sigma_{ij} \equiv (\sigma_i + \sigma_j)/2$.

A total of 100 systems were randomly initialized in a square box at area fraction $\phi = 0.6$, and then isotropically compressed until the system had instantaneous pressure $p > 0.02$.
We chose this pressure threshold because it resulted in a final area fraction $\phi \approx 0.94$ for each system, well beyond the jamming transition and similar to the criterion used in a related study \cite{Manning2011}. 
Pressure was calculated through the virial expression:
\begin{align*}
    p &= \frac{1}{2A} \left[ \frac{1}{2} \sum_{k \neq l} \vec{f}_{kl} \cdot \vec{r}_{kl} \right] .
\end{align*}
In the expression above, $A$ is the total area of the box, $\vec{f}_{kl}$ is the force on particle $k$ due to particle $l$, and $\vec{r}_{kl} = \vec{r}_k - \vec{r}_l$ where $\vec{r}_k$ is the position of particle $k$. 
To compress each system, we followed a protocol (consistent with those of related studies \cite{Gao2006,Chen2018}) of small compression steps until the pressure condition was met.
In each step, we isotropically rescaled each box length such that the area fraction increased by increment $\Delta \phi = 10^{-4}$; then we minimized the energy of the system using the FIRE energy minimization method \cite{Bitzek2006} with maximum step size $dt = 0.005$ within the NVE ensemble.
Convergence of the energy during each minimization step was checked every 100 time steps, and established when both the root-mean-squared force per dimension on each particle and the change in energy per particle dropped below the following thresholds:
\begin{align*}
    \sqrt{\frac{\sum_i f_i^2}{N_{dof}}} <& f_{tol}\\
    \Delta \frac{\sum_i E_i}{N} <& E_{tol}~.
\end{align*}
Above, $f_i$ is the net force on particle $i$, $E_i$ is the total potential energy of particle $i$, and $N_{dof} = dN$, where $d=2$ is the dimensionality of the system and $N$ is the number of particles.
The thresholds for compression were chosen to be $f_{tol} = 10^{-7}$ and $E_{tol} = 10^{-10}$.

Systems were then slowly sheared following an athermal, quasistatic shearing protocol \cite{Malandro1999,Maloney2006} until the final strain $\gamma_f = 0.01$ was reached. 
Each shear step consisted of increasing the strain by an increment of $\Delta \gamma = 10^{-5}$ (similar to that used in related studies \cite{Manning2011, Richard2020a}) and affinely scaling particle positions accordingly. 
Lees-Edwards periodic boundary conditions were used.
After each shear step, energy was minimized using the FIRE energy minimization method \cite{Bitzek2006} with maximum step size $dt = 0.005$ within the NVE ensemble. 
Convergence of the energy was checked every 100 time steps and established when the root-mean-squared force per dimension on each particle and the change in energy per particle dropped below the thresholds $f_{tol} = 10^{-8}$ and $E_{tol} = 10^{-10}$, respectively.
We chose a more stringent force threshold for the shearing than that used during compression to most accurately capture the onset of rearrangement under quasistatic shear.
System snapshots were saved after every energy minimization.

All simulations were performed using the software package HOOMD-blue \cite{Anderson2020}.
During the shearing process, we monitored several indicators of rearrangement, as described in the next section.

\subsection{Quantifying rearrangements}

During the quasistatic shearing of each system, shear stress $\sigma_{xy}$ grows linearly with strain $\gamma$, and abruptly drops during rearrangement events (\textbf{Fig. 1a}, first panel).
Shear stress was calculated via the virial expression:
\begin{align*}
    \sigma_{xy} &= -\frac{1}{2A} \sum_{k \neq l} r_{kl,x}f_{kl,y}~ . 
\end{align*}
In the expression above, $A$ is the area of the simulation box, $r_{kl,x}$ is the $x-$component of the vector $\vec{r}_{kl} = \vec{r}_k - \vec{r}_l$ where $\vec{r}_k$ is the position of particle $k$, and $f_{kl,y}$ is the $y-$component of the vector $\vec{f}_{kl}$ (the force on particle $k$ due to particle $l$).
We found that occasionally rearrangements result in an abrupt rise of $\sigma_{xy}$, rather than a drop.
These strain-hardening events have been observed elsewhere in rapidly quenched systems during athermal quasistatic shearing \cite{Fan2017a}.
We also note that of the 100 jammed systems generated, 38 resulted in an initial negative shear stress (\textbf{Fig. S1}), indicating some initial instability of the packing with respect to shear. 
Similarly, we observed that rearrangements in our systems sometimes resulted in packings transitioning from positive shear stress to negative shear stress.
It was previously found that this phenomenon can occur for finite jammed systems which were prepared via isotropic compression \cite{Dagois-Bohy2012,Goodrich2014b}.
To ameliorate this issue, we ultimately only chose to analyze rearrangements for which $\sigma_{xy} \geq 0$ at the start of the rearrangement.
Although 7 of the 100 rearrangements chosen in this way still result in negative shear stress afterward, our restriction means that systems are not in an anomalous, unstable regime at the onset of the rearrangements we analyze.

The rearrangement event can be quantified in several other ways, which we review here.
Like the shear stress, the potential energy of the system also indicates rearrangement; it grows with strain $\gamma$, and abruptly drops during rearrangement events (\textbf{Fig. 1a}, second panel).
The rearrangement event can also be quantified using the per-particle metric $D^2_{i,min}(t,t+1)$ \cite{Falk1998}, the mean squared difference between the actual displacements of particle $i$'s neighbors over some time interval and the corresponding displacements from the best-fit affine deformation of the neighborhood:
\begin{align*}
    D^2_{i,min}(t,t+1) &= \frac{1}{N} \sum_{j\in nn} \left[ \vec{r}_{ji}(t+1) - J_{min} \vec{r}_{ji} (t) \right]^2~.
\end{align*}
In the expression above, $j$ indexes over all $N$ neighbors of particle $i$, $\vec{r}_{ji} = \vec{r}_j - \vec{r}_i$, and $J_{min}$ is the best-fit local affine deformation tensor around particle $i$ that minimizes $D^2_{i,min}(t,t+1)$.
We chose to define the neighborhood of each particle $i$ as all particles $j$ for which $r_{ji} < 2$.
Details regarding the calculation of $J_{min}$ for each particle can be found in the \emph{Supplementary Information}. 
The quantity $D^2_{i,min}(t,t+1)$ characterizes the extent to which the local neighborhood around particle $i$ evolves in a non-affine manner between times $t$ and $t+1$, and thus it indicates rearrangement around particle $i$ between those times. 
The average $\langle D^2_{min}(t,t+1) \rangle$, taken over all particles in the system, spikes during rearrangement events (\textbf{Fig. 1a}, third panel), with a small subset of particles contributing the majority of that average (\textbf{Fig. 1b}, right panel). 
In general, larger $\langle D^2_{min}(t,t+1) \rangle$ corresponds to a larger drop in potential energy, $\Delta PE (t,t+1) = PE(t+1) - PE(t)$ (\textbf{Fig. S2}).

The rearrangement event can be alternately quantified by examining the time evolution of the eigenvalues and eigenvectors of the dynamical matrix $D$ of the system: 
$D_{ij} \equiv \left. \frac{\partial^2 U}{\partial x_{i}\partial x_{j}} \right\vert_0$, where $U$ is the potential energy of the system, $x_i$ is its $i$-th degree of freedom, and the Hessian is evaluated about some stable equilibrium configuration $\vec{x}_0$. 
Rattlers in the system, defined as particles with less than three neighbor contacts according to Maxwell constraint counting, are entirely removed from the dynamical matrix prior to each diagonalization.
Once the rattlers are removed, each dynamical matrix has two remaining trivial zero-frequency eigenmodes, corresponding to global translations of the system.
Diagonalization is performed using the \emph{eigsh} method in the python package \emph{scipy.sparse.linalg}, which is designed to save time and space by working efficiently with sparse matrices; when necessary for numerical precision, a full singular value decomposition was used via the \emph{svd} method in the python package \emph{scipy.linalg}.
Rearrangements in the system correspond to the drop of one eigenvalue to 0 (\textbf{Fig. 1a}, fourth panel). 
This eigenvalue is associated with an eigenvector, termed the critical mode, which represents the zero-energy vibrational mode of the system immediately prior to rearrangement. 
Particles that participate most in the critical mode thus represent the locus of rearrangement (\textbf{Fig. 1b}, left panel).
Participation in the critical mode can be identified via large $\vert e_{i,C} \vert^2$, where $\vec{e}_{i,C}$ is the polarization vector of particle $i$ in the critical mode. 
The entries of $\vec{e}_{i,C}$ are the values of the critical mode for all degrees of freedom associated with particle $i$, and 0 for all other degrees of freedom.
We stress that the locus of rearrangement is not necessarily the only particle set that rearranges during a system-wide rearrangement event, however: subsequent particle rearrangements might also occur during the energy minimization process.
These rearrangements would have high values of $D^2_{min}$ but would not participate heavily in the critical mode.

For each simulation, we choose as our rearrangement event the pair of frames corresponding to times $(t,t+1)$ for which $\Delta PE(t,t+1)$ is maximal (and for which $\sigma_{xy}(t) > 0$, as mentioned above). 
This subset of rearrangement events has a lower correlation in general between the locus of rearrangement and which particles rearrange most according to $D^2_{min}$ (\textbf{Fig. S3}). 
Our choice enables us to investigate the predictive capability of controllability for a broad range of rearrangement events, including those for which rearrangement is generally confined to the critical mode (\textbf{Fig. 1b}) and those for which rearrangement generally occurs in avalanche-like events beyond the critical mode (\textbf{Fig. 1c}).

\begin{figure}
    \centering
    \includegraphics[width=\textwidth]{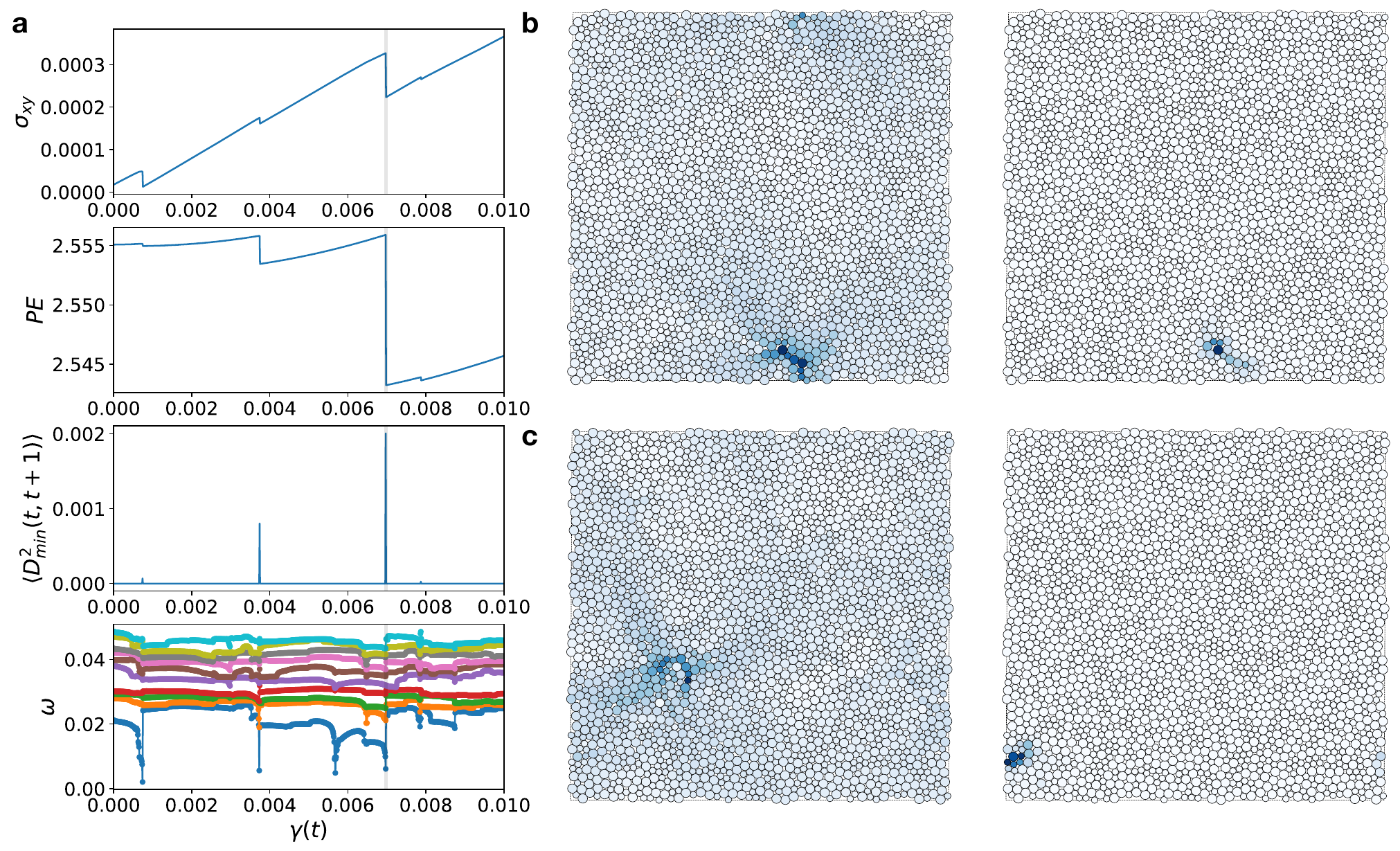}
    \caption{
    \textbf{Rearrangements can be quantified in a jammed disordered solid undergoing quasistatic shear.}
    \textbf{(a)} Shear stress $\sigma_{xy}$, potential energy $PE$, average non-affine motion $\langle D^2_{min} \rangle$, and the eigenfrequencies $\omega$ of the dynamical matrix vary as a function of strain $\gamma (t)$. 
    Rearrangement events are characterized by drops in $\sigma_{xy}$, $PE$, and the critical eigenfrequency of the dynamical matrix, and a spike in $\langle D^2_{min} \rangle$. 
    By choosing the rearrangement corresponding to the largest drop in $PE$ in each simulation, indicated with a transparent black line in panel (a), we select for a set of rearrangement events for which the critical mode (panels b and c, left) can range from high correlation with the $D^2_{min}$ map of rearrangements (panel b, right) to low correlation with the $D^2_{min}$ map of rearrangements (panel c, right). 
    \textbf{(b)} Snapshot of a system prior to its rearrangement event. 
    This rearrangement event has the highest Spearman correlation between the critical mode (left) and the $D^2_{min}$ map (right). 
    \textbf{(c)} Snapshot of another system prior to its rearrangement event. 
    This rearrangement event has the lowest Spearman correlation between the critical mode (left) and the $D^2_{min}$ map (right).
    Each Spearman correlation is calculated only for the particles with $D^2_{min}$ values in the top 95th percentile of all values in the snapshot, in order to limit the calculation to the particle subset that moves most during rearrangement and thereby avoid noise in the correlation. 
    In each panel, particles in the leftmost image are colored linearly according to the magnitude of their polarization vector in the critical mode, with blue corresponding to the highest magnitude.  
    Particles in the rightmost panel are colored linearly according to $D^2_{min}$, with blue corresponding to the highest value.
    }
    \label{fig:fig1}
\end{figure}

\subsection{Controllability}
We next sketch the basics of control theory in the context of a jammed solid and introduce average controllability.
We begin by letting $N$ be the number of particles in our system, and $d$ be its dimension, such that there are $dN$ configurational degrees of freedom.
If the system is mechanically stable at some equilibrium configuration $(x_1^0,x_2^0,\dots x_{dN}^0)$, we can take a harmonic approximation to find its potential energy for small perturbations $(x_1,x_2,\dots x_{dN})$ around this equilibrium configuration:

\begin{align*}
    U(x_1,x_2,\dots,x_{dN}) &\approx U_0 + \frac{1}{2} \sum_{i=1}^{dN} \sum_{j=1}^{dN} \left. \frac{\partial^2 U}{\partial x_{i} \partial x_{j}} \right\vert_0 x_{i} x_{j}~.
\end{align*}
In the expression above, $U_0$ is the potential energy evaluated at the equilibrium configuration $(x_1^0,x_2^0,\dots x_{dN}^0)$, the partial derivative is also evaluated at the equilibrium configuration, and $x_{i}$ is the $i$-th configurational degree of freedom in the system.

The force on the $k$-th degree of freedom is then:
\begin{align*}
    F_{k} &= - \frac{\partial U}{\partial x_{k}} \\
    &= - \sum_{i=1}^{dN} \left. \frac{\partial^2 U}{\partial x_{i}\partial x_{k}} \right\vert_0 x_{i}~.
\end{align*}
And thus we can establish the dynamics of each configurational degree of freedom:
\begin{align*}
    m_k \ddot{x}_{k} &= - \sum_{i=1}^{dN} \left. \frac{\partial^2 U}{\partial x_{i}\partial x_{k}} \right\vert_0 x_{i}~,
\end{align*}
where $m_k$ is the mass of the particle associated with the degree of freedom $x_k$. 
In matrix form:
\begin{align*}
    M \ddot{X} &= - D X ~,
\end{align*}
where $X$ is a vector of configurational degrees of freedom $x_{i}$, $D$ is the dynamical matrix defined by $D_{ij} \equiv \left. \frac{\partial^2 U}{\partial x_{i}\partial x_{j}} \right\vert_0$, and $M$ encodes mass via $M_{ij} = m_i \delta_{ij}$ where $\delta_{ij}$ is the Kronecker delta.

We can re-write as a linear matrix expression:
\begin{align*}
    \dot{Y} &= A Y ~,
\end{align*}
where $Y \equiv \begin{bmatrix} X \\ \dot{X} \end{bmatrix}$ and $A \equiv \begin{bmatrix} 0 & I \\ -M^{-1}D & 0 \end{bmatrix} $.
If external force $F_{ext,i}$ is applied to each degree of freedom $x_i$, then this expression becomes:
\begin{align*}
    \dot{Y} &= A Y + B u(t) ~,
\end{align*}
where $B \equiv \begin{bmatrix} 0 & 0 \\ 0 & M^{-1} \end{bmatrix} $ and $u(t) \equiv \begin{bmatrix} 0 \\ F_{ext} \end{bmatrix}$.
$F_{ext}$ is a vector of forces acting on each degree of freedom in the system.

The general solution of the above equation at any time $T$, given known input forces and initial state $Y(0)$, is the convolution of the input with the homogeneous response of the system:
\begin{align*}
    Y(T) &= Y(0) e^{AT} + \int_0^T e^{A(T-t)} B u(t) dt ~.
\end{align*}

The average controllability of the system is the strength of the impulse response of the system given input forces to particles that are specified by $B$. 
It is the trace of the controllability Gramian, $Tr(W_C)$, where:
\begin{align*}
    W_C &= \int_0^T e^{At} B B^T e^{A^Tt} dt~.
\end{align*}
Above, $B^T$ and $A^T$ are the transposes of the matrices $B$ and $A$, respectively.
In the following, we will set all masses equal to 1 such that $M=I$.
Then:
\begin{align*}
    e^{At} = \begin{bmatrix} \cos \sqrt{D}t & \frac{1}{\sqrt{D}}\sin \sqrt{D}t \\ -\sqrt{D}\sin \sqrt{D}t &  \cos \sqrt{D}t \end{bmatrix}~,
\end{align*}
and the full control ($B= \begin{bmatrix} 0 & 0 \\ 0 & I \end{bmatrix}$, such that all degrees of freedom are capable of receiving control input) Gramian is:

\begin{align*}
    W_C 
&= \begin{bmatrix} \frac{-1}{4\sqrt{D^3}} \left[ \sin 2 \sqrt{D}T - 2 \sqrt{D}T \right] & \frac{1}{2D} \sin^2 \sqrt{D}T \\ \frac{1}{2D} \sin^2 \sqrt{D}T & \frac{1}{4 \sqrt{D}} \left[ \sin 2 \sqrt{D}T + 2 \sqrt{D}T \right] \end{bmatrix}~.
\end{align*}
Then average controllability is:
\begin{align*}
    Tr(W_C) &= Tr \left( \frac{-1}{4\sqrt{D^3}} \left[ \sin 2 \sqrt{D}T - 2 \sqrt{D}T \right] \right) + Tr \left( \frac{1}{4 \sqrt{D}} \left[ \sin 2 \sqrt{D}T + 2 \sqrt{D}T \right] \right) \\
    &= Tr \left[ f(D) \right]~,
\end{align*}
where the function $f(D)$ is defined as
\begin{align*}
f(D) \equiv \frac{-1}{4\sqrt{D^3}} \left[ \sin 2 \sqrt{D}T - 2 \sqrt{D}T \right] + \frac{1}{4 \sqrt{D}} \left[ \sin 2 \sqrt{D}T + 2 \sqrt{D}T \right]~.
\end{align*}
Since the trace of any square matrix is a sum of its eigenvalues, it is useful to find the eigenvalues of $f(D)$ to efficiently calculate the average controllability.
By the spectral mapping theorem \cite{Dunford1988}, the eigenvalues of $f(D)$ are equal to $f(\lambda)$, the function applied to the eigenvalues $\lambda$ of $D$. 
In other words, the eigenvalue set of $f(D)$ is $\{ \frac{-1}{4\sqrt{\lambda_k^3}} \left[ \sin 2 \sqrt{\lambda_k}T - 2 \sqrt{\lambda_k}T \right] + \frac{1}{4 \sqrt{\lambda_k}} \left[ \sin 2 \sqrt{\lambda_k}T + 2 \sqrt{\lambda_k}T \right] \}$, where $\lambda_k$ is the $k$-th eigenvalue of $D$.
Average controllability is the sum of these eigenvalues.
For clarity, we will define $\lambda_k \equiv \omega_k^2$.
The quantity $\omega_k$ is the frequency of the $k$-th eigenmode of $D$.
The trace (after simplifying) is thus:

\begin{align*}
    Tr(W_C) &= \sum_k f(\omega_k^2) \\
    &= \sum_{k} \left[ \frac{T}{2\omega_k^2} (\omega_k^2 + 1) + \frac{1}{4\omega_k^3} \sin (2\omega_k T) (\omega_k^2 - 1) \right]~.
\end{align*}

\noindent
Note that, for the spectral mapping theorem to apply in the above derivation, the function $f$ must be analytic on a neighborhood of the spectrum of $D$ \cite{Dunford1988}.
$D$ has at least two zero eigenvalues (corresponding to the global translational degrees of freedom of the system).
However, $f$ is analytic around the value 0 for any finite $T$, so the spectral mapping theorem applies.
To see this, note that $\sin (2 x^\frac{1}{2} T) = 2 x^\frac{1}{2} T - \frac{(2 x^\frac{1}{2} T)^3}{6} + O(x^\frac{5}{2})$.
Thus, $f(x) = T + \frac{T^3}{3} + O(x)$.
In the neighborhood of $x = 0$, this function is therefore analytic. 

\begin{figure}
    \centering
    \includegraphics[width=\textwidth]{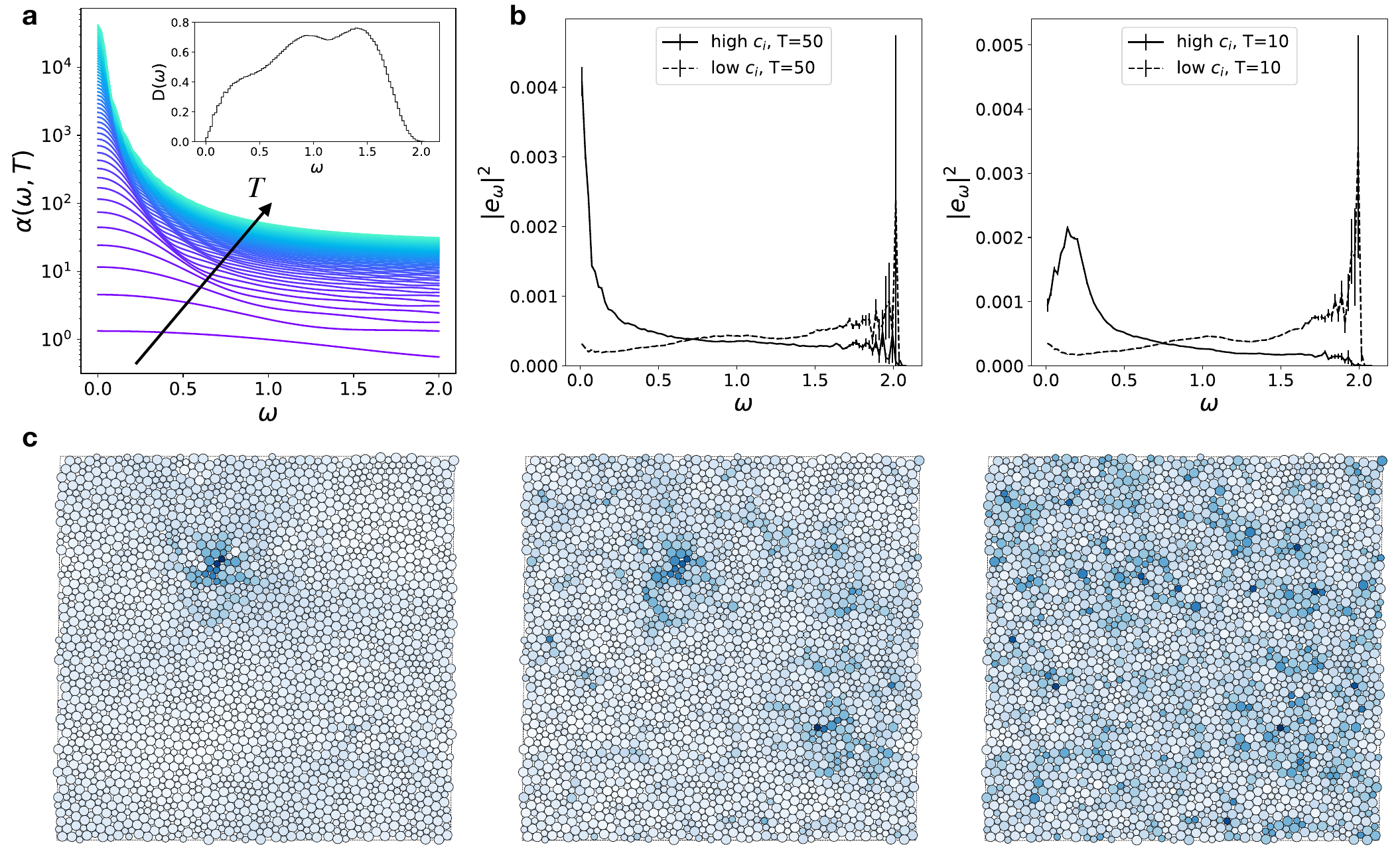}
    \caption{
    \textbf{High-controllability particles participate in varying distributions of eigenmodes in a way that depends on the associated time horizon.}
    \textbf{(a)} The prefactor $\alpha$ (defined in the main text) as a function of $\omega$, for varying values of the time horizon $T$.
    Values of $T$ range from 1 (the most purple curve) to 50 (the most green curve).
    As $T$ increases, the prefactor weights low-frequency eigenmodes more and more heavily with respect to high-frequency eigenmodes.
    Inset: The density of states of the system immediately prior to rearrangement.
    Values of $\omega$ were aggregated over all frames immediately prior to rearrangement for all state points.
    \textbf{(b)} The squared magnitude of the polarization vector ($\vert e_\omega \vert^2$) of high- (solid line) and low- (dotted line) controllability particles in all eigenmodes indexed by $\omega$, for $T=50$ (left) and $T=10$ (right).
    High-controllability particles at the longer time horizon $T=50$ participate more heavily in the lowest-frequency eigenmodes.
    Each curve is an average taken over the 10 particles with the highest or lowest average controllability for all systems at the frame immediately prior to rearrangement. 
    Values of $\omega$ were grouped into 100 bins, and averages were taken over all polarization vectors associated with each bin.
    Error bars are the standard deviation of the mean.
    \textbf{(c)} Snapshots of an example system immediately prior to its rearrangement event, colored by the critical mode (left), average controllability at $T=50$ (middle), and average controllability at $T=10$ (right).
    At the longer time horizon $T=50$, the map of average controllability resembles the critical mode map.
    Particles in the left image are colored linearly according to the magnitude of their polarization vector in the critical mode, with blue corresponding to the highest magnitude.
    Particles in the middle and right images are colored logarithmically according to their average controllability, with blue corresponding to the maximum value in each snapshot.
    }
    \label{fig:fig2}
\end{figure}

If external force can only be applied to particle $i$, the average controllability of the system (or the average controllability of particle $i$, $c_i$) is:

\begin{align*}
    c_i &= \sum_k \left[ \frac{T}{2\omega_k^2} (\omega_k^2 + 1) + \frac{1}{4\omega_k^3} \sin (2\omega_k T) (\omega_k^2 - 1) \right] \vert e_{i,k} \vert^2\\
    &\equiv \sum_k \alpha (\omega_k, T) \vert e_{i,k} \vert^2~.
\end{align*}

\noindent where the prefactor $\alpha (\omega_k, T)$ is identical for all particles and depends only on the eigenfrequencies $\omega_k$ and the time horizon $T$, and $\vec{e}_{i,k}$ is the polarization vector of particle $i$ in eigenmode $k$: its entries are the values of the eigenmode for all degrees of freedom associated with particle $i$, and 0 otherwise.
The derivation of $c_i$ is contained in the \emph{Supplemental Information}.
Average controllability for each particle is thus a weighted sum over its participation in each eigenmode, where the weight is given by the prefactor $\alpha (\omega, T)$.
This sum is not evenly distributed across values of $\omega$, but instead distributed according to the density of states of the system (\textbf{Fig. \ref{fig:fig2}a}, inset), with most modes corresponding to intermediate values of $\omega$ between 0.5 and 1.5, and fewer low- and high-frequency modes.
We note that, when calculating $c_i$ in practice, we set $\alpha (\omega_k, T) = T + \frac{T^3}{3}$ for all eigenmodes for which $\vert \omega_k^2 \vert < 10^{-7}$, to avoid numerical precision issues.

The weighting behavior of the prefactor $\alpha$ changes according to the time horizon $T$, and thus shifts which particles are high vs. low controllability.
We can generally consider two regimes for the behavior of $\alpha(\omega, T)$.
The first occurs when $\omega T < 1$, and the second occurs for $\omega T > 1$.
First, we consider the regime for which $\omega T < 1$, or $\omega < \frac{1}{T}$.
In this regime, we can consider the first few terms of the Taylor expansion of $\sin (2\omega T)$ to find the dependence of $\alpha(\omega, T)$ on $\omega$ and $T$ when $\omega < \frac{1}{T}$:
\begin{align*}
    \alpha(\omega,T) &= \frac{T}{2\omega^2} (\omega^2 + 1) + \frac{1}{4\omega^3} \sin (2\omega T) (\omega^2 - 1)\\
	&\approx \frac{T}{2\omega^2} (\omega^2 + 1) + \frac{1}{4\omega^3}  \left[ 2 \omega T - \frac{(2\omega T)^3}{6} + \frac{(2\omega T)^5}{120} + O(\omega T)^7 \right](\omega^2 - 1)\\
    &\approx T + \frac{T^3}{3} -\omega^2 \left( \frac{T^3}{3} + \frac{T^5}{15} \right) + O(\omega^4T^5)~.
\end{align*}
The above implies that $\alpha(\omega,T)$ grows as $T$ grows, with a value given by $T + \frac{T^3}{3}$ at $\omega = 0$.
Values of $\alpha(\omega,T)$ near $\omega = 0$ decrease according to the curvature $\frac{T^3}{3} + \frac{T^5}{15}$; thus $\alpha(\omega,T)$ exhibits a sharper and sharper drop-off with increasing $\omega$ as $T$ grows.
These two behaviors can be seen in the plot of $\alpha (\omega, T)$ as a function of $\omega$ (\textbf{Fig. 2a}): The y-intercept of each curve and the negative slope of each curve near $\omega=0$ grow in magnitude as the curves become more green, corresponding to increasing values of $T$.

In the second regime, when $\omega T > 1$ and thus $\omega > \frac{1}{T}$, we cannot consider only the first few terms in the Taylor expansion of $\sin (2\omega T)$, and we must examine the full expression for $\alpha(\omega, T)$ in order to understand its behavior:
\begin{align*}
    \alpha(\omega,T) &= \frac{T}{2\omega^2} (\omega^2 + 1) + \frac{1}{4\omega^3} \sin (2\omega T) (\omega^2 - 1)\\
    &= \frac{T}{2} + \frac{1}{4\omega} \sin (2\omega T) + \frac{T}{2\omega^2} - \frac{1}{4\omega^3} \sin (2\omega T)\\
    &\approx \frac{T}{2} + \frac{T}{2\omega^2}~.
\end{align*}
The approximation in the final line could be taken because $\vert \sin (2\omega T) \leq 1\vert$, and $\omega > \frac{1}{T}$ in this regime (or equivalently, $T > \frac{1}{\omega}$).
Thus, $\vert \frac{T}{2} \vert > \vert \frac{1}{4\omega} \sin (2\omega T) \vert$, and $\vert \frac{T}{2\omega^2} \vert > \vert \frac{1}{4\omega^3} \sin (2\omega T) \vert$.
In other words, the two terms that scale with $T$ dominate the two terms that do not scale with $T$, and we can approximate $\alpha(\omega,T) \approx \frac{T}{2} + \frac{T}{2\omega^2}$ especially for $\omega T \gg 1$.
As $\omega$ grows, $\alpha(\omega,T) \rightarrow \frac{T}{2}$.
This behavior can be seen in the plot of $\alpha (\omega, T)$ as a function of $\omega$ (\textbf{Fig. 2a}):  
At larger $\omega$, the curves approach a baseline value that increases as the curves become more green, corresponding to increasing values of $T$.

The cross-over between the regimes occurs at $\omega T \sim 1$.
As $T$ grows, this cross-over occurs at smaller and smaller values of $\omega$, and $\alpha(\omega,T)$ becomes a curve that is very sharply peaked around $\omega = 0$.
Particles with the highest values of controllability at long $T$ thus participate heavily in the lowest-frequency eigenmodes (Fig. S4), since $\alpha$ weights these modes heavily in the average controllability sum (\textbf{Fig. 2b}, left).
The map of high-controllability particles at long $T$ is similar to the map of particles that participate most in the critical mode (\textbf{Fig. 2c}, left and middle).
At smaller $T$, the cross-over occurs at larger and larger values of $\omega$, and $\alpha(\omega,T)$ is much less sharply peaked around $\omega=0$.
This can be seen in the plot of $\alpha (\omega, T)$ as a function of $\omega$ (\textbf{Fig. 2a}): For the lowest value of $T = 1$, corresponding to the most purple curve, $\alpha$ is close to a flat line.
Particles with the highest values of controllability at short $T$ thus participate in higher energy eigenmodes, since $\alpha$ weights these modes more equally in the average controllability sum (\textbf{Fig. 2b}, right), and there are many more eigenmodes at these higher frequencies (\textbf{Fig. 2a}, inset).
The map of high-controllability particles at short $T$ is dissimilar to the critical mode map (\textbf{Fig. 2c}, right).

Note that the low-controllability particles shown in \textbf{Fig. 2b} and (to a lesser extent) the high-controllability particles shown in \textbf{Figs. 2b} and \textbf{S4} also have significant participation in the highest-energy eigenmodes (albeit with large standard deviation across particles).
Since there are very few of these highest-energy eigenmodes (\textbf{Fig. 2a}, inset), this participation does not significantly influence the particles' average controllability.

\subsection{Vibrality}
To evaluate average controllability in the context of other known and successful metrics for predicting particle rearrangements, we calculate the vibrality \cite{Tong2014} of all particles in our systems.
Prior work has demonstrated that vibrality successfully predicts rearrangement in a variety of contexts \cite{Richard2020a}.
The vibrality $v_i$ of particle $i$ is defined as:

\begin{align*}
    v_i &\equiv \sum_j \frac{\vert e_{i,j} \vert^2}{\omega_j^2} ~.
\end{align*}

\noindent The sum is over the eigenmodes of the dynamical matrix, each of which has frequency $\omega_j$.
As defined in the previous section, $\vec{e}_{i,j}$ is the polarization vector of particle $i$ in eigenmode $j$.
Note that we exclude the two eigenmodes with the smallest eigenvalues, corresponding to the two trivial zero-frequency global translations of the system, from the above sum over eigenmodes.

This metric, also grounded in linear response and the diagonalization of the dynamical matrix, is (like average controllability) a weighted sum over the particle's participation in each eigenmode.
However, in contrast to average controllability, the weights do not depend on any variable other than the frequency of each eigenmode $j$; they are set at $\frac{1}{\omega_j^2}$.
Thus vibrality is higher for particles that participate more in low-frequency modes.
By contrast, average controllability is tuned by the time horizon $T$: At long $T$, average controllability is also higher for particles that participate more in low-frequency modes.
At short $T$, average controllability is higher for particles that participate more in higher-frequency modes.
In fact, average controllability is directly related to vibrality: It can be written as the scaled vibrality of a particle plus corrective terms that depend on the time horizon $T$ (see the \emph{Supplementary Information} for more detail.
These corrective terms enable average controllability, unlike vibrality, to encode the time-dependent nature of the system's response to external input.

\subsection{Quantifying predictive capability}
To quantify the capability of average controllability to predict which particles participate in rearrangement events, we calculate $c_i$ for each particle over a set of snapshots preceding the rearrangement, and examine if particles with high values of $c_i$ at these earlier snapshots are those that ultimately rearrange most. 
We follow a similar procedure to that described in Ref. \citenum{Richard2020a}, in order to most closely compare our controllability metric to those evaluated in that paper.

If each rearrangement occurs at strain $\gamma_r$, then each snapshot at earlier strains $\gamma_p$ can be characterized by $\Delta \gamma \equiv \gamma_r - \gamma_p$, the amount of strain the system subsequently undergoes until rearrangement. 
The snapshots we consider range from $\Delta \gamma = 0$, or the snapshot immediately prior to the rearrangement event, to $\Delta \gamma = 0.002$, which approximates the average strain interval $\langle \Delta \gamma \rangle = 2.11 \times 10^{-3}$ (taken over all simulations) between each selected rearrangement event and the immediately prior non-negligible rearrangement event.
We define a non-negligible rearrangement event as any in which potential energy decreases from one frame to the next and has difference $\Delta PE (t,t+1) > 1.1 \times 10^{-5}$.
For rearrangements with values of $\Delta PE$ above the threshold $1.1 \times 10^{-5}$, $\log \langle D^2_{min} \rangle$, or the logarithm of the average value of $D^2_{min}$ over all particles during the rearrangement, scales linearly with $\log \Delta PE$.
By contrast, rearrangements with values of $\Delta PE$ below $1.1\times 10^{-5}$ correspond to a wide range of $\langle D^2_{min} \rangle$ values, including a large cluster ranging from $10^{-8}$ to $10^{-10}$, that do not follow the same scaling with $\Delta PE$.
For this reason, we interpret rearrangements with values of $\Delta PE < 1.1\times 10^{-5}$ as negligible.

For each snapshot, we calculate $c_i$ for each particle for a range of time horizons at all integers between $T=1$ and $T=100$, and evaluate each associated rank $r_{c,i}$ of $c_i$.
The upper limit $T=100$ is well above $1/\langle \omega_C \rangle$, where $\langle \omega_C \rangle = 0.015$ is the average frequency of the critical mode taken over all selected rearrangement events.
The rank is normalized such that $r_{c,i} = 1$ for the particle with the highest value of $c_i$ in each snapshot, and $r_{c,i} = 0$ for the particle with the lowest value of $c_i$ in each snapshot.
We then track $\langle r_{c} \rangle$ for snapshots of increasing $\Delta \gamma$ for the particles that rearrange most during the rearrangement event, where the average is taken over all systems.

We perform an identical analysis using vibrality, calculating $\langle r_{v} \rangle$ for the particles that rearrange most over snapshots of increasing $\Delta \gamma$.
Thus we are able to directly compare the predictive capacity of average controllability to that of vibrality.

\section{Results}

We first examine the rearrangement prediction capabilities of average controllability at set values of the time horizon $T$.
We find that at the long time horizon $T = 50$, controllability is a successful predictor of particle rearrangement, on par with vibrality.
The normalized rank of controllability $\langle r_c \rangle$ for the particle that participates most in the critical mode at rearrangement remains above 0.875 for all $\Delta \gamma \leq 0.002$, indicating that the most rearranging particle is in the top 15\% of controllability at frames well before the actual rearrangement event (\textbf{Fig. 3a}).
This behavior is nearly identical for $\langle r_v \rangle$, the normalized rank of vibrality for the particle that participates most in the critical mode.
$\langle r_c \rangle$ remains above 0.8 for all $\Delta \gamma <= 0.002$ even for particles that participate 2nd-, 3rd-, 4th-, and 5th-most in the critical mode at rearrangement (\textbf{Fig. S5}).
This behavior is also matched by that of vibrality, indicating that both controllability and vibrality are highly correlated with multiple particles in the rearrangement locus in our systems, even well before the rearrangement event.

\begin{figure}
    \centering
    \includegraphics[width=0.8\textwidth]{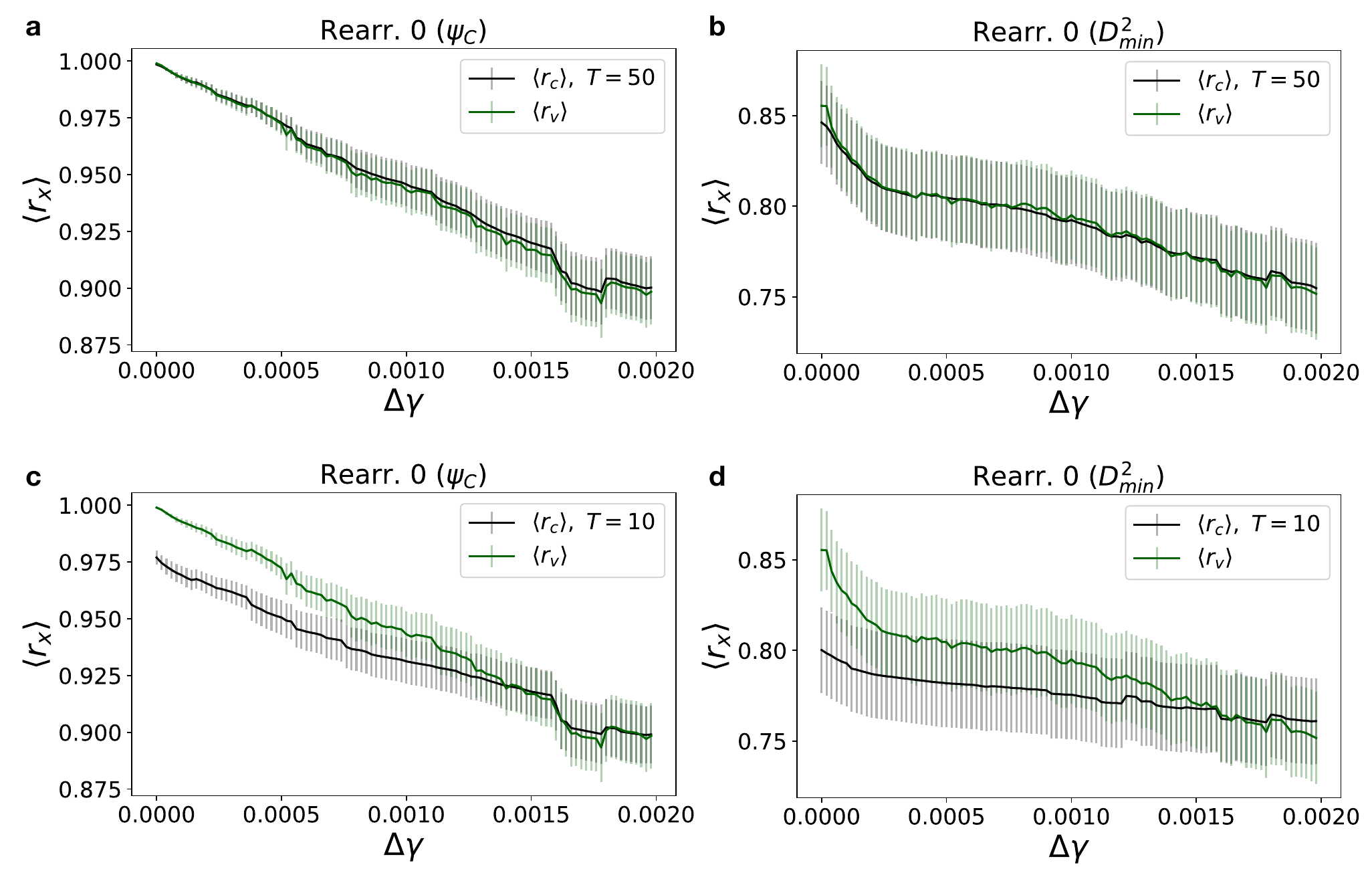}
    \caption{
    \textbf{Average controllability at long $T$ is as successful a predictor of rearrangement as vibrality, and average controllability at short $T$ hints at the eigenmode participation of rearrangers.}
    \textbf{(a,b)} Normalized rank of average controllability at $T = 50$ ($\langle r_c \rangle$, black) and vibrality ($\langle r_v \rangle$, green) as a function of $\Delta \gamma$ for the particles that participate most in the critical mode at rearrangement (panel a) and have the highest value of $D^2_{min}$ during rearrangement (panel b). 
    $\langle r_c \rangle$ for $T=50$ is as high as $\langle r_v \rangle$ in all cases.
    \textbf{(c,d)} Normalized rank of average controllability at $T = 10$ ($\langle r_c \rangle$, black) and vibrality ($\langle r_v \rangle$, green) as a function of $\Delta \gamma$ for the particles that participate most in the critical mode at rearrangement (panel c) and have the highest value of $D^2_{min}$ during rearrangement (panel d).
    $\langle r_c \rangle$ for $T=10$ is lower than $\langle r_v \rangle$ for a range of $\Delta \gamma$, and approximately equal to or greater than $\langle r_v \rangle$ for $\Delta \gamma \gtrsim 0.0015$, indicating that the rearranging particles participate in higher energy modes further from the rearrangement event on average.
    For all curves, averages are taken over all simulations, and error bars indicate standard error of the mean.
    }
    \label{fig:fig3}
\end{figure}

We also measure particle rearrangement in an alternative way, according to non-affine motion during the rearrangement event as captured by $D^2_{min}$, to test the predictive capacity of our metric in extended avalanche-like rearrangement scenarios.
We find that average controllability at $T=50$ continues to be a successful predictor of which particles rearrange according to $D^2_{min}$, although it is not as successful in this context as the previous one in which rearrangement was defined according to the critical mode.
The normalized rank of controllability $\langle r_c \rangle$ for the particle with the highest value of $D^2_{min}$ during the rearrangement event is between 0.85 and 0.75 for all $\Delta \gamma \leq 0.002$ (\textbf{Fig. 3b}).
As with rearrangement defined according to the critical mode, vibrality follows an almost identical pattern to average controllability at $T=50$.
This is unsurprising since average controllability at long $T$ and vibrality are both high for similar types of particles (those that participate most in low-frequency eigenmodes).
That the rearranging particles according to $D^2_{min}$ would have lower values of average controllability and vibrality at all $\Delta \gamma$ is also unsurprising, since these particles are more likely to participate in higher-frequency eigenmodes.
Both average controllability and vibrality are similarly successful in predicting which particles have the 2nd-, 3rd-, 4th-, and 5th-highest values of $D^2_{min}$ during the rearrangement event (\textbf{Fig. S9}).

At long time horizons, our controllability metric is equivalent to vibrality in terms of its predictive capacity; we next examine the predictive capabilities of controllability at a shorter time horizon in order to better understand the physics of the system.
For example, if we consider the time horizon $T=10$, we find that average controllability is a less successful predictor than vibrality at $\Delta \gamma < 0.0015$, but as successful as or more successful than vibrality at $\Delta \gamma > 0.0015$.
The normalized rank $\langle r_c \rangle < \langle r_v \rangle$ for $\Delta \gamma < 0.0015$ both when the top rearranger is defined according to its participation in the critical mode (\textbf{Fig. 3c}) and its value of $D^2_{min}$ (\textbf{Fig. 3d}).
In both cases, values of $\langle r_c \rangle$ are still relatively high, and $\langle r_c \rangle \gtrsim \langle r_v \rangle$ for $\Delta \gamma > 0.0015$.
These results indicate that rearranging particles according to both metrics participate in higher energy modes further from the rearrangement event, since average controllability at the shorter time horizon $T=10$ (indicating participation in higher-frequency eigenmodes) improves in prediction accuracy with respect to vibrality further from the rearrangement event in both cases.
For the particles that participate 2nd-, 3rd-, 4th-, and 5th-most in the critical mode at rearrangement, we note that $\langle r_c \rangle$ is progressively lower than its value for the top rearranger for all $\Delta \gamma$ (\textbf{Fig. S6}), whereas for the particles with the 2nd-, 3rd-, 4th-, and 5th-highest values of $D^2_{min}$ during rearrangement, $\langle r_c \rangle$ is uniformly lower than its value for the top rearranger for all $\Delta \gamma$ (\textbf{Fig. S10}).
These results indicate that the weighting into eigenmodes captured by average controllability at short $T$ is not as suitable for predicting which particles rearrange 2nd, 3rd, 4th, and 5th most during rearrangement events.

We next take further advantage of the parameter $T$ and the flexible weighting into eigenmodes that it affords to gain additional physical insight into the system.
Instead of using average controllability at a fixed value for all systems and all values of $\Delta \gamma$, we examine results when we choose the value of $T$ that optimizes controllability of the rearranging particle at each snapshot, for each system.
Uncovering the optimal $T$ for each rearranging particle gives insight into which distributions of eigenmodes are especially predictive of rearrangement in each context.
Additionally, using this protocol significantly improves the predictive capability of average controllability beyond that of vibrality, for the particle that participates most in the critical mode at rearrangement (\textbf{Fig. 4a}), the particle with the highest value of $D^2_{min}$ during rearrangement (\textbf{Fig. 4b}), and the 2nd-, 3rd-, 4th-, and 5-th top rearrangers according to both the critical mode and $D^2_{min}$ (\textbf{Figs. S7} and \textbf{S11}).
For all rearranging particles according to the critical mode, $\langle r_c \rangle$ calculated using the optimal $T$ is significantly above $\langle r_v \rangle$ at all $\Delta \gamma$, with values at $\Delta \gamma = 0.002$ ranging from $\approx 0.95$ (for the top rearranger) to $\approx 0.875$ (for the 5th-highest rearranger).
For all rearranging particles according to $D^2_{min}$, $\langle r_c \rangle$ calculated using the optimal $T$ is significantly above $\langle r_v \rangle$ at all $\Delta \gamma$, with values at $\Delta \gamma = 0.002$ ranging from $\approx 0.86$ (for the top rearranger) to $\approx 0.825$ (for the 5th-highest rearranger).

The values of optimal $T$ as a function of $\Delta \gamma$, for each rearrangement context, are especially informative.
The quantity $\langle T_{optimal} \rangle$ decreases as a function of $\Delta \gamma$ for the particle that participates most in the critical mode at rearrangement (\textbf{Fig. 4c}), the particle with the highest value of $D^2_{min}$ during rearrangement (\textbf{Fig. 4d}), and the 2nd-, 3rd-, 4th-, and 5-th top rearrangers according to both the critical mode and $D^2_{min}$ (\textbf{Figs. S8} and \textbf{S12}).
The results provide additional evidence that every rearranging particle, no matter how rearrangement is defined, participates in higher energy eigenmodes at earlier times before the rearrangement event.
Put another way, these rearranging particles participate in lower energy vibrational modes on approach to rearrangement. 
That every type of rearranging particle we studied showcases this phenomenon, even according to two separate definitions of rearrangement, could be explained by the fact that the relevant vibrational modes themselves decrease in energy approaching rearrangement. 
On any elastic branch between rearrangements, the critical mode and other low energy modes decrease in frequency \cite{Maloney2006}.

Furthermore, $\langle T_{optimal} \rangle$ for all rearranging particles according to $D^2_{min}$ is generally lower than $\langle T_{optimal} \rangle$ for all rearranging particles according to the critical mode, indicating that the rearranging particles with high values of $D^2_{min}$ participate in higher energy eigenmodes than their critical mode counterparts at all values of $\Delta \gamma$.

\begin{figure}
    \centering
    \includegraphics[width=0.8\textwidth]{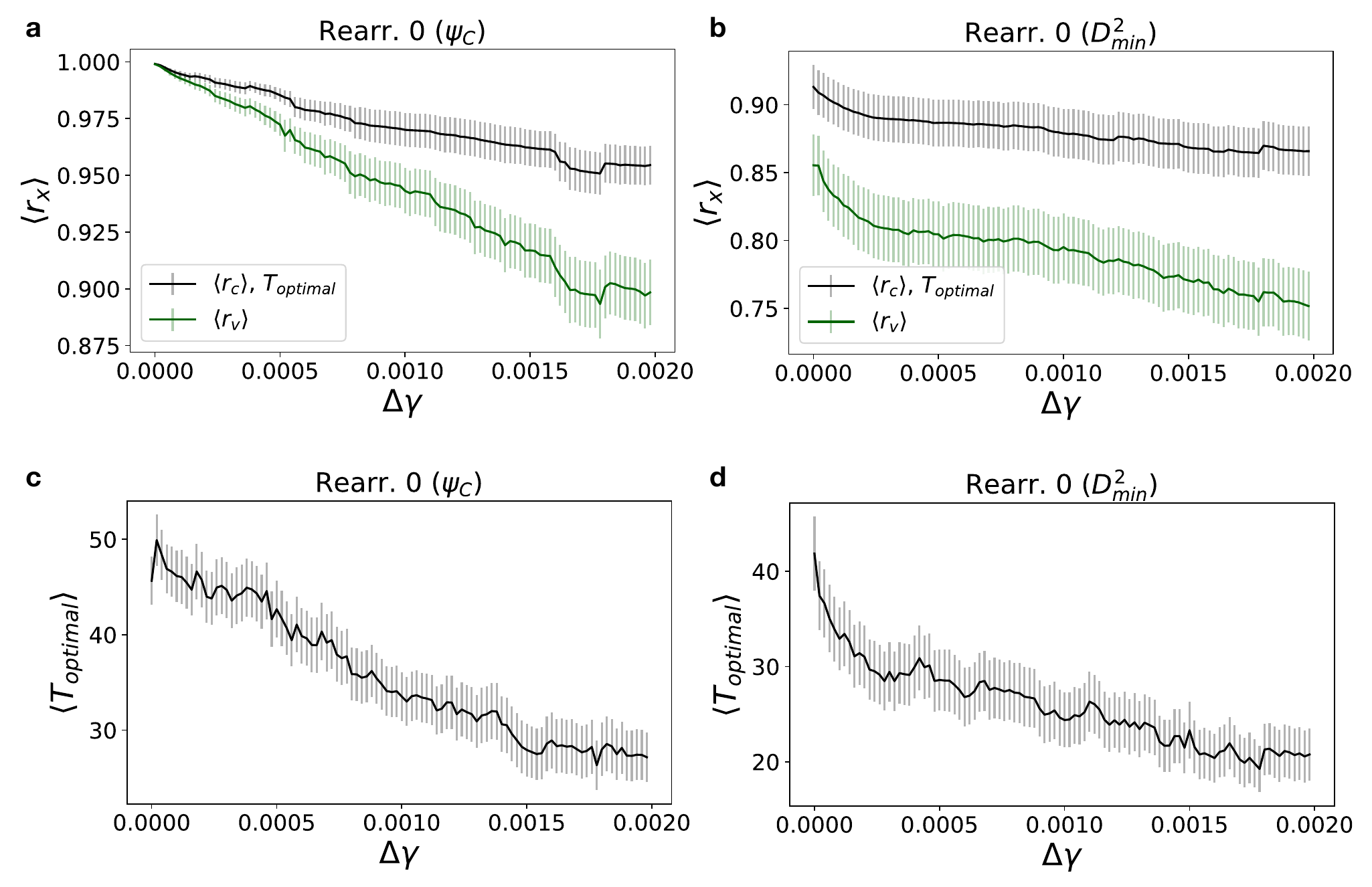}
    \caption{
    \textbf{Choosing $T$ to maximize the average controllability of the rearranging particle at each system snapshot provides insight into the eigenmode participation of rearrangers.}
    \textbf{(a,b)} Normalized rank of average controllability at the time horizon value $T_{optimal}$ that maximizes it at each snapshot of each system ($\langle r_c \rangle$, black) and vibrality ($\langle r_v \rangle$, green) of the rearranging particle as a function of $\Delta \gamma$. 
    Plots are shown for two definitions of rearranging particles: those that participate most in the critical mode at rearrangement (panel a) and those that have the highest value of $D^2_{min}$ during rearrangement (panel b). 
    $\langle r_c \rangle$ is significantly higher than $\langle r_v \rangle$ in all cases.
    \textbf{(c,d)} The value $T_{optimal}$ used to calculate each normalized rank as a function of $\Delta \gamma$ for the particles that participate most in the critical mode at rearrangement (panel c) and have the highest value of $D^2_{min}$ during rearrangement (panel d).
    $\langle T_{optimal} \rangle$ decreases in both cases as a function of $\Delta \gamma$, indicating that the rearranging particles participate in higher energy modes further from the rearrangement event on average.
    Additionally, $\langle T_{optimal} \rangle$ for the rearrangers determined according to $D^2_{min}$ is lower in general than $\langle T_{optimal} \rangle$ for the rearrangers determined according to the critical mode, indicating that the rearrangers determined according to $D^2_{min}$ participate in higher energy modes in general.
    For all curves, averages are taken over all simulations, and error bars are standard error of the mean.
    }
    \label{fig:fig4}
\end{figure}

\section{Discussion}

With this study, we sought to introduce linear control in the context of jammed disordered materials, and demonstrate its usefulness and efficacy in the prediction of rearranging particles under quasistatic shear.
We derived an expression for average controllability per particle as a weighted sum over the participation of the particle in each eigenmode of the dynamical matrix, and showed that this weighted sum varies with the time horizon associated with controllability.
Through the correlation of controllability with rearrangement, we demonstrated that the time horizon is a tunable parameter that provides insight into which eigenmodes matter for rearrangement in various contexts.
We found that average controllability at a long time horizon, which is highest for particles that participate most in the lowest-frequency eigenmodes of the dynamical matrix, is a strong predictor of rearrangement according to the critical mode of the system at rearrangement.
Its predictive capability is on par with that of vibrality, a well-established metric for predicting rearrangement that is also higher for particles that participate more in low-frequency modes.
Average controllability at a long time horizon, like vibrality, is less predictive of rearrangement according to a more general quantity, $D^2_{min}$, that captures total non-affine motion during the rearrangement event (and likely involves higher energy modes in the system).

Our investigation of average controllability over a range of time horizon values demonstrates the power of concepts from control theory to provide physical insight into material systems.
We maximized average controllability's predictive capacity by choosing time horizons to optimize it for each system and each snapshot prior to rearrangement.
Using this protocol, we increased the predictive capacity of average controllability well beyond that of vibrality, and additionally verified through examining the optimal time horizon that rearranging particles according to both the critical mode and $D^2_{min}$ tend to participate in higher energy eigenmodes of the dynamical matrix further from the rearrangement event.

Future extensions of our method could move beyond the single-particle level by calculating average controllability for subsets of particles simultaneously receiving external force. 
Such a metric would enable the investigation of multi-particle rearrangements. 
Another extension would involve the calculation of average controllability in the case that particles only receive external force along one dimension, enabling a closer examination of rearrangement as a function of perturbation direction.
A final extension could investigate rearrangements in scenarios in which external perturbations occur over a finite time interval, through careful tuning of the time horizon $T$ when calculating average controllability.

Control theory has a rich history of application for designing the properties of complex systems \cite{Liu2016}.
For example, many studies have examined how to modify the topology of theoretical networks in order to design their controllability (the ability to be driven to an arbitrary final state) \cite{Lvlin2013, Chanekar2019}, structural controllability (controllability under some reweighting of the network edges) \cite{Chen2019}, controllability under a minimal number of nodes that receive control input \cite{Wang2012b}, and the robustness of controllability against attacks \cite{Xiao2014}.
Beyond these theoretical studies, the control theory framework has been used to design collective behavior in robot flocks \cite{Mateo2019}, the navigation of model animals \cite{Mori2025} and synthetic active particles \cite{Yang2018a} through complex environments, the mitigation of virus spread throughout a population \cite{Roy2009}, the self-assembly of colloidal systems \cite{Tang2022}, and the behavior of synthetic genetic circuits in living cells \cite{DelVecchio2016}. 
An increasing number of researchers have drawn connections between control theory, statistical physics \cite{DSouza2023}, and soft matter \cite{Alvarado2025}. 
Within the field of soft matter, however, few studies have utilized control theory to design mechanical response in disordered granular materials \cite{Alvarado2025}.
Our study, through the establishment of the control framework in jammed granular materials and the demonstration of its usefulness for understanding particle rearrangement in these systems, motivates future work at the intersection of control and amorphous materials.

For example, control theory might be useful in the design of allosteric response in jammed materials \cite{Rocks2017, Kim2019, Stern2021, Dillavou2022}, in which external force applied to a specified subset of particles causes a localized mechanical response in a separate subset of particles.
Control theoretic techniques are already widely applied to identify which node subsets of networks must receive control input to optimize various network properties \cite{Liu2011, Fitch2016, Summers2014a, Li2015, DU2024141605} including the controllability of a subset of target nodes \cite{Gao2014,Klickstein2017}, making control an ideal framework to apply to this problem.
For example, one could calculate the external control input or set of particles receiving that input that is required to facilitate that mechanical response with minimal control energy (or magnitude of the control input integrated over time \cite{Lindmark2018}).
This ``minimum energy control" problem is a subtype of the more general optimal control problem \cite{Sargent2000} that has been investigated extensively for decades in engineering (and, more recently, physics \cite{Alvarado2025}).
Using the tools developed for this problem, one could even design material configurations that optimize the minimal control energy required to facilitate allosteric response, as explored recently in models of neural systems \cite{Kim2018a}.
Control theory might also be used to determine which dynamical patterns of external forcing cause specific failure modes in materials in various contexts, for the purposes of designing materials that propagate stress in directed ways, or fail in a manner that is controllable through specified strain protocols.
This method of reverse-engineering control input to achieve a desired or observed dynamical response has recently proven fruitful to determine which regions of the brain provide input during cognitive tasks \cite{Liang2024}.
Finally, control theory may be useful for the design of global material properties like elastic and shear moduli.
Previous efforts to design these properties have focused on the selective removal of bonds in systems \cite{Goodrich2015,Hexner2018a}, controlling the boundaries of systems during aging \cite{Pashine2019a}, and periodic driving of systems during aging \cite{Hexner2020}.
Control theory may be helpful for this design problem by establishing a theoretical framework for how particle interactions and material structure influence dynamical response under control input.

In order to control finite particle displacements in these materials, techniques beyond linear control theory will be useful, including generic nonlinear control \cite{Zanudo2017,Rozum2022}, techniques to control complementarity systems which are often used to model contact mechanics in robotics \cite{Brogliato2003,Hogan2022}, and the inclusion of damping in the dynamics when appropriate.
Data-driven techniques \cite{Baggio2021}, or techniques accounting for input constraints \cite{Cornelius2013}, may also be useful when applying control to more realistic and experimental systems. 
Future work may also be extended to material systems beyond those studied here, including systems subject to thermal perturbations, systems of particles with anisotropic interactions, and elastic networks which may be useful models for proteins or other macromolecules \cite{Flechsig2017,Flechsig2018}.

\section{Citation diversity statement}
Recent work has identified a bias in citation practices such that papers from women and other marginalized scholars in STEM are under-cited relative to expected rates. 
Here we sought to proactively consider choosing references that reflect the diversity of the field in thought, form of contribution, gender, and other factors. 
We use databases that store the probability of a name being carried by people of different genders to mitigate our own citation bias at the intersection of name and identity. 
By this measure (and excluding self-citations to the first and last authors of our current paper, and papers whose authors' first names could not be determined), our references contain 4.08\% woman(first)/woman(last), 13.27\% man/woman, 10.20\% woman/man, and 72.45\% man/man categorization. 
This method is limited in that names, pronouns, and social media profiles used to construct the databases may not, in every case, be indicative of gender identity. 
Furthermore, probabilistic studies of names cannot be used to detect citation costs that are specific to intersex, non-binary, or transgender people who are out to a large number of their colleagues. 
We look forward to future work that could help us to better understand how to support equitable practices in science.

\section{Acknowledgments}
E.G.T. acknowledges support by the International Human Frontier Science Program Organization (HFSPO) under grant RGEC33/2024.
J.Z.K. acknowledges support by postdoctoral fellowships from the Bethe/KIC/Wilkins, Mong Neurotech, and the Eric and Wendy Schmidt AI in Science program of Schmidt Sciences, LLC.
D.S.B. acknowledges support from the Center for Curiosity.

All authors conceptualized the work and reviewed and edited the manuscript. 
E.G.T performed the simulations and analysis and wrote the manuscript.

\bibliographystyle{apsrev4-2}
\bibliography{control}

@book{Dunford1988,
  title = {Linear {{Operators}}. {{Part}} 1: {{General Theory}}},
  shorttitle = {Linear {{Operators}}. {{Part}} 1},
  author = {Dunford, Nelson and Schwartz, Jacob T.},
  year = 1988,
  publisher = {Interscience Publishers},
  address = {New York},
  isbn = {978-0-470-22605-6},
  langid = {english}
}

@article{Cornelius2013,
  title = {Realistic Control of Network Dynamics},
  author = {Cornelius, Sean P. and Kath, William L. and Motter, Adilson E.},
  year = 2013,
  month = jun,
  journal = {Nature Communications},
  volume = {4},
  number = {1},
  pages = {1942},
  publisher = {Nature Publishing Group},
  issn = {2041-1723},
  doi = {10.1038/ncomms2939},
  urldate = {2026-02-11},
  copyright = {2013 Springer Nature Limited},
  langid = {english}
}

@article{Baggio2021,
  title = {Data-Driven Control of Complex Networks},
  author = {Baggio, Giacomo and Bassett, Danielle S. and Pasqualetti, Fabio},
  year = 2021,
  month = mar,
  journal = {Nature Communications},
  volume = {12},
  number = {1},
  pages = {1429},
  publisher = {Nature Publishing Group},
  issn = {2041-1723},
  doi = {10.1038/s41467-021-21554-0},
  urldate = {2026-02-11},
  copyright = {2021 The Author(s)},
  langid = {english}
}

@article{Rozum2022,
  title = {Leveraging Network Structure in Nonlinear Control},
  author = {Rozum, Jordan and Albert, R{\'e}ka},
  year = 2022,
  month = oct,
  journal = {npj Systems Biology and Applications},
  volume = {8},
  number = {1},
  pages = {36},
  publisher = {Nature Publishing Group},
  issn = {2056-7189},
  doi = {10.1038/s41540-022-00249-2},
  urldate = {2026-02-11},
  copyright = {2022 The Author(s)},
  langid = {english}
}

@article{Zanudo2017,
  title = {Structure-Based Control of Complex Networks with Nonlinear Dynamics},
  author = {Za{\~n}udo, Jorge Gomez Tejeda and Yang, Gang and Albert, R{\'e}ka},
  year = 2017,
  month = jul,
  journal = {Proceedings of the National Academy of Sciences},
  volume = {114},
  number = {28},
  pages = {7234--7239},
  publisher = {Proceedings of the National Academy of Sciences},
  doi = {10.1073/pnas.1617387114},
  urldate = {2026-02-11}
}

@article{Hexner2020,
  title = {Periodic Training of Creeping Solids},
  author = {Hexner, Daniel and Liu, Andrea J. and Nagel, Sidney R.},
  year = 2020,
  month = dec,
  journal = {Proceedings of the National Academy of Sciences},
  volume = {117},
  number = {50},
  pages = {31690--31695},
  publisher = {Proceedings of the National Academy of Sciences},
  doi = {10.1073/pnas.1922847117},
  urldate = {2026-02-10}
}

@article{Pashine2019a,
  title = {Directed Aging, Memory, and Nature's Greed},
  author = {Pashine, Nidhi and Hexner, Daniel and Liu, Andrea J. and Nagel, Sidney R.},
  year = 2019,
  journal = {Science Advances},
  volume = {5},
  number = {12},
  eprint = {1903.05776},
  pages = {1--8},
  issn = {23752548},
  doi = {10.1126/sciadv.aax4215},
  archiveprefix = {arXiv}
}

@article{Hexner2018a,
  title = {Role of Local Response in Manipulating the Elastic Properties of Disordered Solids by Bond Removal},
  author = {Hexner, Daniel and Liu, Andrea J. and Nagel, Sidney R.},
  year = 2018,
  journal = {Soft Matter},
  volume = {14},
  number = {2},
  eprint = {1708.07207},
  pages = {312--318},
  publisher = {Royal Society of Chemistry},
  issn = {17446848},
  doi = {10.1039/c7sm01727h},
  archiveprefix = {arXiv},
  pmid = {29251303}
}

@article{Goodrich2015,
  title = {The {{Principle}} of {{Independent Bond-Level Response}}: {{Tuning}} by {{Pruning}} to {{Exploit Disorder}} for {{Global Behavior}}},
  author = {Goodrich, Carl P. and Liu, Andrea J. and Nagel, Sidney R.},
  year = 2015,
  journal = {Physical Review Letters},
  volume = {114},
  number = {22},
  eprint = {1502.02953},
  pages = {1--5},
  issn = {10797114},
  doi = {10.1103/PhysRevLett.114.225501},
  archiveprefix = {arXiv},
  pmid = {26196627}
}

@misc{Liang2024,
  title = {Reverse Engineering the Brain Input: {{Network}} Control Theory to Identify Cognitive Task-Related Control Nodes},
  shorttitle = {Reverse Engineering the Brain Input},
  author = {Liang, Zhichao and Zhang, Yinuo and Wu, Jushen and Liu, Quanying},
  year = 2024,
  month = apr,
  number = {arXiv:2404.16357},
  eprint = {2404.16357},
  primaryclass = {q-bio},
  publisher = {arXiv},
  doi = {10.48550/arXiv.2404.16357},
  urldate = {2026-02-10},
  archiveprefix = {arXiv}
}

@article{Kim2018a,
  title = {Role of Graph Architecture in Controlling Dynamical Networks with Applications to Neural Systems},
  author = {Kim, Jason Z. and Soffer, Jonathan M. and Kahn, Ari E. and Vettel, Jean M. and Pasqualetti, Fabio and Bassett, Danielle S.},
  year = 2018,
  journal = {Nature Physics},
  volume = {14},
  number = {1},
  eprint = {1702.00354},
  pages = {91--98},
  issn = {17452481},
  doi = {10.1038/NPHYS4268},
  archiveprefix = {arXiv},
  isbn = {1073858416667},
  pmid = {29422941}
}

@article{Sargent2000,
  title = {Optimal Control},
  author = {Sargent, Roger W. H.},
  year = 2000,
  month = dec,
  journal = {Journal of Computational and Applied Mathematics},
  series = {Numerical {{Analysis}} 2000. {{Vol}}. {{IV}}: {{Optimization}} and {{Nonlinear Equations}}},
  volume = {124},
  number = {1},
  pages = {361--371},
  issn = {0377-0427},
  doi = {10.1016/S0377-0427(00)00418-0},
  urldate = {2026-02-11}
}

@article{Lindmark2018,
  title = {Minimum Energy Control for Complex Networks},
  author = {Lindmark, Gustav and Altafini, Claudio},
  year = 2018,
  month = feb,
  journal = {Scientific Reports},
  volume = {8},
  number = {1},
  pages = {3188},
  publisher = {Nature Publishing Group},
  issn = {2045-2322},
  doi = {10.1038/s41598-018-21398-7},
  urldate = {2026-02-11},
  copyright = {2018 The Author(s)},
  langid = {english}
}

@article{Klickstein2017,
  title = {Energy Scaling of Targeted Optimal Control of Complex Networks},
  author = {Klickstein, Isaac and Shirin, Afroza and Sorrentino, Francesco},
  year = 2017,
  month = apr,
  journal = {Nature Communications},
  volume = {8},
  number = {1},
  pages = {15145},
  publisher = {Nature Publishing Group},
  issn = {2041-1723},
  doi = {10.1038/ncomms15145},
  urldate = {2026-02-11},
  copyright = {2017 The Author(s)},
  langid = {english}
}

@article{Gao2014,
  title = {Target Control of Complex Networks},
  author = {Gao, Jianxi and Liu, Yang-Yu and D'Souza, Raissa M. and Barab{\'a}si, Albert-L{\'a}szl{\'o}},
  year = 2014,
  month = nov,
  journal = {Nature Communications},
  volume = {5},
  number = {1},
  pages = {5415},
  publisher = {Nature Publishing Group},
  issn = {2041-1723},
  doi = {10.1038/ncomms6415},
  urldate = {2026-02-11},
  copyright = {2014 The Author(s)},
  langid = {english}
}

@article{DU2024141605,
  title = {Evaluation of Key Node Groups of Embodied Carbon Emission Transfer Network in {{China}} Based on Complex Network Control Theory},
  author = {Du, Ruijin and Zhang, Mengxi and Zhang, Nidan and Liu, Yue and Dong, Gaogao and Tian, Lixin and Kong, Ziyang and Ahsan, Muhammad},
  year = 2024,
  journal = {Journal of Cleaner Production},
  volume = {448},
  pages = {141605},
  issn = {0959-6526},
  doi = {10.1016/j.jclepro.2024.141605}
}

@article{Li2015,
  title = {Minimum-Cost Control of Complex Networks},
  author = {Li, Guoqi and Hu, Wuhua and Xiao, Gaoxi and Deng, Lei and Tang, Pei and Pei, Jing and Shi, Luping},
  year = 2015,
  month = dec,
  journal = {New Journal of Physics},
  volume = {18},
  number = {1},
  pages = {013012},
  publisher = {IOP Publishing},
  issn = {1367-2630},
  doi = {10.1088/1367-2630/18/1/013012},
  urldate = {2026-02-11},
  langid = {english}
}

@article{Dillavou2022,
  title = {Demonstration of {{Decentralized Physics-Driven Learning}}},
  author = {Dillavou, Sam and Stern, Menachem and Liu, Andrea J. and Durian, Douglas J.},
  year = 2022,
  month = jul,
  journal = {Physical Review Applied},
  volume = {18},
  number = {1},
  pages = {014040},
  publisher = {American Physical Society},
  doi = {10.1103/PhysRevApplied.18.014040},
  urldate = {2026-02-10}
}

@article{Stern2021,
  title = {Supervised {{Learning}} in {{Physical Networks}}: {{From Machine Learning}} to {{Learning Machines}}},
  shorttitle = {Supervised {{Learning}} in {{Physical Networks}}},
  author = {Stern, Menachem and Hexner, Daniel and Rocks, Jason W. and Liu, Andrea J.},
  year = 2021,
  month = may,
  journal = {Physical Review X},
  volume = {11},
  number = {2},
  pages = {021045},
  publisher = {American Physical Society},
  doi = {10.1103/PhysRevX.11.021045},
  urldate = {2026-02-10}
}

@misc{Alvarado2025,
  title = {Optimal {{Control}} in {{Soft}} and {{Active Matter}}},
  author = {Alvarado, Jos{\'e} and Teich, Erin and Sivak, David and Bechhoefer, John},
  year = 2025,
  month = apr,
  number = {arXiv:2504.08676},
  eprint = {2504.08676},
  primaryclass = {cond-mat},
  publisher = {arXiv},
  doi = {10.48550/arXiv.2504.08676},
  urldate = {2025-05-29},
  archiveprefix = {arXiv}
}

@article{DSouza2023,
  title = {Controlling Complex Networks with Complex Nodes},
  author = {D'Souza, Raissa M. and Di Bernardo, Mario and Liu, Yang-Yu},
  year = 2023,
  month = mar,
  journal = {Nature Reviews Physics},
  volume = {5},
  number = {4},
  pages = {250--262},
  issn = {2522-5820},
  doi = {10.1038/s42254-023-00566-3},
  urldate = {2025-01-24},
  langid = {english}
}

@article{DelVecchio2016,
  title = {Control Theory Meets Synthetic Biology},
  author = {Del Vecchio, Domitilla and Dy, Aaron J. and Qian, Yili},
  year = 2016,
  month = jul,
  journal = {Journal of The Royal Society Interface},
  volume = {13},
  number = {120},
  pages = {20160380},
  issn = {1742-5689},
  doi = {10.1098/rsif.2016.0380},
  urldate = {2026-02-10}
}

@article{Tang2022,
  title = {Control of {{Microparticle Assembly}}},
  author = {Tang, Xun and Grover, Martha A.},
  year = 2022,
  month = may,
  journal = {Annual Review of Control, Robotics, and Autonomous Systems},
  volume = {5},
  number = {1},
  pages = {491--514},
  issn = {2573-5144, 2573-5144},
  doi = {10.1146/annurev-control-042920-100621},
  urldate = {2025-02-21},
  langid = {english}
}

@inproceedings{Roy2009,
  title = {A Network Control Theory Approach to Virus Spread Mitigation},
  booktitle = {2009 {{IEEE Conference}} on {{Technologies}} for {{Homeland Security}}},
  author = {Roy, Sandip and Wan, Yan and Saberi, Ali},
  year = 2009,
  month = may,
  pages = {599--606},
  doi = {10.1109/THS.2009.5168092},
  urldate = {2026-02-10}
}

@article{Yang2018a,
  title = {Optimal {{Navigation}} of {{Self-Propelled Colloids}}},
  author = {Yang, Yuguang and Bevan, Michael A.},
  year = 2018,
  month = nov,
  journal = {ACS Nano},
  volume = {12},
  number = {11},
  pages = {10712--10724},
  publisher = {American Chemical Society},
  issn = {1936-0851},
  doi = {10.1021/acsnano.8b05371},
  urldate = {2026-02-10}
}

@article{Mori2025,
  title = {Optimal Switching Strategies for Navigation in Stochastic Settings},
  author = {Mori, Francesco and Mahadevan, L.},
  year = 2025,
  month = jun,
  journal = {Journal of The Royal Society Interface},
  volume = {22},
  number = {227},
  pages = {20240677},
  issn = {1742-5689},
  doi = {10.1098/rsif.2024.0677},
  urldate = {2026-03-05}
}

@article{Mateo2019,
  title = {Optimal Network Topology for Responsive Collective Behavior},
  author = {Mateo, David and Horsevad, Nikolaj and Hassani, Vahid and Chamanbaz, Mohammadreza and Bouffanais, Roland},
  year = 2019,
  month = apr,
  journal = {Science Advances},
  volume = {5},
  number = {4},
  pages = {eaau0999},
  publisher = {American Association for the Advancement of Science},
  doi = {10.1126/sciadv.aau0999},
  urldate = {2026-02-10}
}

@article{Xiao2014,
  title = {Optimization of Robustness of Network Controllability against Malicious Attacks},
  author = {{Xiao, Yan-Dong} and {Lao, Song-Yang} and {Lv-Lin}, Hou and Bai, Liang},
  year = 2014,
  month = sep,
  journal = {Chinese Physics B},
  volume = {23},
  number = {11},
  pages = {118902},
  issn = {1674-1056},
  doi = {10.1088/1674-1056/23/11/118902},
  urldate = {2026-02-10},
  langid = {english}
}

@article{Wang2012b,
  title = {Optimizing Controllability of Complex Networks by Minimum Structural Perturbations},
  author = {Wang, Wen-Xu and Ni, Xuan and Lai, Ying-Cheng and Grebogi, Celso},
  year = 2012,
  month = feb,
  journal = {Physical Review E},
  volume = {85},
  number = {2},
  pages = {026115},
  publisher = {American Physical Society},
  doi = {10.1103/PhysRevE.85.026115},
  urldate = {2026-02-10}
}

@article{Chen2019,
  title = {Minimal {{Edge Addition}} for {{Network Controllability}}},
  author = {Chen, Ximing and Pequito, S{\'e}rgio and Pappas, George J. and Preciado, Victor M.},
  year = 2019,
  month = mar,
  journal = {IEEE Transactions on Control of Network Systems},
  volume = {6},
  number = {1},
  pages = {312--323},
  issn = {2325-5870},
  doi = {10.1109/TCNS.2018.2814841},
  urldate = {2026-01-23}
}

@inproceedings{Chanekar2019,
  title = {Network {{Modification}} Using a {{Novel Gramian-based Edge Centrality}}},
  booktitle = {2019 {{IEEE}} 58th {{Conference}} on {{Decision}} and {{Control}} ({{CDC}})},
  author = {Chanekar, Prasad Vilas and Nozari, Erfan and Cortes, Jorge},
  year = 2019,
  month = dec,
  pages = {1686--1691},
  publisher = {IEEE},
  address = {Nice, France},
  doi = {10.1109/CDC40024.2019.9028860},
  urldate = {2026-02-10},
  copyright = {https://ieeexplore.ieee.org/Xplorehelp/downloads/license-information/IEEE.html},
  isbn = {978-1-7281-1398-2},
  langid = {english}
}

@inproceedings{Lvlin2013,
  title = {Enhancing {{Complex Network Controllability}} by {{Rewiring Links}}},
  booktitle = {2013 {{Third International Conference}} on {{Intelligent System Design}} and {{Engineering Applications}}},
  author = {Lvlin, Hou and Songyang, Lao and Jiang, Bu and Liang, Bai},
  year = 2013,
  month = jan,
  pages = {709--711},
  doi = {10.1109/ISDEA.2012.168},
  urldate = {2026-02-10}
}

@article{Liu2016,
  title = {Control Principles of Complex Systems},
  author = {Liu, Yang-Yu and Barab{\'a}si, Albert-L{\'a}szl{\'o}},
  year = 2016,
  month = sep,
  journal = {Reviews of Modern Physics},
  volume = {88},
  number = {3},
  pages = {035006},
  issn = {0034-6861, 1539-0756},
  doi = {10.1103/RevModPhys.88.035006},
  urldate = {2025-01-24},
  copyright = {http://link.aps.org/licenses/aps-default-license},
  langid = {english}
}

@misc{Bianchin2018,
  title = {Gramian-{{Based Optimization}} for the {{Analysis}} and {{Control}} of {{Traffic Networks}}},
  author = {Bianchin, Gianluca and Pasqualetti, Fabio},
  year = 2018,
  month = nov,
  number = {arXiv:1811.02673},
  eprint = {1811.02673},
  primaryclass = {math},
  publisher = {arXiv},
  doi = {10.48550/arXiv.1811.02673},
  urldate = {2026-01-23},
  archiveprefix = {arXiv}
}

@inproceedings{Chapman2011,
  title = {{{UAV}} Flocking with Wind Gusts: {{Adaptive}} Topology and Model Reduction},
  shorttitle = {{{UAV}} Flocking with Wind Gusts},
  booktitle = {Proceedings of the 2011 {{American Control Conference}}},
  author = {Chapman, Airlie and Mesbahi, Mehran},
  year = 2011,
  month = jun,
  pages = {1045--1050},
  publisher = {IEEE},
  address = {San Francisco, CA},
  doi = {10.1109/ACC.2011.5990799},
  urldate = {2026-01-23},
  isbn = {978-1-4577-0081-1 978-1-4577-0080-4 978-1-4577-0079-8},
  langid = {english}
}

@article{Guzman2020,
  title = {Topology Optimization of Piezoelectric Sensor and Actuator Layers for Active Vibration Control},
  author = {Guzm{\'a}n, Daniel Giraldo and Silva, Emilio C N and Rubio, Wilfredo Montealegre},
  year = 2020,
  month = jun,
  journal = {Smart Materials and Structures},
  volume = {29},
  number = {8},
  pages = {085009},
  publisher = {IOP Publishing},
  issn = {0964-1726},
  doi = {10.1088/1361-665X/ab9061},
  urldate = {2026-01-23},
  langid = {english}
}

@article{GonzalezVillasanti2019,
  title = {A {{Control-Theoretic Assessment}} of {{Interventions During Drinking Events}}},
  author = {Gonzalez Villasanti, Hugo and Passino, Kevin M. and Clapp, John D. and Madden, Danielle R.},
  year = 2019,
  month = feb,
  journal = {IEEE Transactions on Cybernetics},
  volume = {49},
  number = {2},
  pages = {604--615},
  issn = {2168-2275},
  doi = {10.1109/TCYB.2017.2782010},
  urldate = {2026-01-23}
}

@article{Summers2014a,
  title = {Optimal {{Sensor}} and {{Actuator Placement}} in {{Complex Dynamical Networks}}},
  author = {Summers, Tyler H. and Lygeros, John},
  year = 2014,
  month = jan,
  journal = {IFAC Proceedings Volumes},
  series = {19th {{IFAC World Congress}}},
  volume = {47},
  number = {3},
  pages = {3784--3789},
  issn = {1474-6670},
  doi = {10.3182/20140824-6-ZA-1003.00226},
  urldate = {2026-01-22}
}

@inproceedings{Fitch2016,
  title = {Optimal Leader Selection for Controllability and Robustness in Multi-Agent Networks},
  booktitle = {2016 {{European Control Conference}} ({{ECC}})},
  author = {Fitch, Katherine and Leonard, Naomi Ehrich},
  year = 2016,
  month = jun,
  pages = {1550--1555},
  publisher = {IEEE},
  address = {Aalborg, Denmark},
  doi = {10.1109/ECC.2016.7810511},
  urldate = {2026-01-22},
  isbn = {978-1-5090-2591-6},
  langid = {english}
}

@article{Parkes2024,
  title = {A Network Control Theory Pipeline for Studying the Dynamics of the Structural Connectome},
  author = {Parkes, Linden and Kim, Jason Z. and Stiso, Jennifer and Brynildsen, Julia K. and Cieslak, Matthew and Covitz, Sydney and Gur, Raquel E. and Gur, Ruben C. and Pasqualetti, Fabio and Shinohara, Russell T. and Zhou, Dale and Satterthwaite, Theodore D. and Bassett, Dani S.},
  year = 2024,
  month = dec,
  journal = {Nature Protocols},
  volume = {19},
  number = {12},
  pages = {3721--3749},
  publisher = {Nature Publishing Group},
  issn = {1750-2799},
  doi = {10.1038/s41596-024-01023-w},
  urldate = {2026-01-22},
  copyright = {2024 Springer Nature Limited},
  langid = {english}
}

@article{Amon2013,
  title = {Experimental Investigation of Plastic Deformations before a Granular Avalanche},
  author = {Amon, Axelle and Bertoni, Roman and Crassous, J{\'e}r{\^o}me},
  year = {2013},
  month = jan,
  journal = {Physical Review E},
  volume = {87},
  number = {1},
  pages = {012204},
  publisher = {American Physical Society},
  doi = {10.1103/PhysRevE.87.012204},
  urldate = {2024-07-19}
}

@article{Anderson2020,
  title = {{{HOOMD-blue}}: {{A Python}} Package for High-Performance Molecular Dynamics and Hard Particle {{Monte Carlo}} Simulations},
  author = {Anderson, Joshua A. and Glaser, Jens and Glotzer, Sharon C.},
  year = {2020},
  journal = {Computational Materials Science},
  volume = {173},
  eprint = {1308.5587},
  pages = {109363},
  publisher = {Elsevier},
  issn = {09270256},
  doi = {10.1016/j.commatsci.2019.109363},
  archiveprefix = {arXiv}
}

@article{Argon1979,
  title = {Plastic Deformation in Metallic Glasses},
  author = {Argon, Ali S},
  year = {1979},
  month = jan,
  journal = {Acta Metallurgica},
  volume = {27},
  number = {1},
  pages = {47--58},
  issn = {0001-6160},
  doi = {10.1016/0001-6160(79)90055-5},
  urldate = {2024-07-18}
}

@article{Bapst2020,
  title = {Unveiling the Predictive Power of Static Structure in Glassy Systems},
  author = {Bapst, Victor and Keck, T. and {Grabska-Barwi{\'n}ska}, A. and Donner, C. and Cubuk, E. D. and Schoenholz, S. S. and Obika, A. and Nelson, A. W.R. and Back, T. and Hassabis, D. and Kohli, Pushmeet},
  year = {2020},
  journal = {Nature Physics},
  volume = {16},
  number = {4},
  pages = {448--454},
  issn = {17452481},
  doi = {10.1038/s41567-020-0842-8}
}

@article{Bechhoefer2005,
  title = {Feedback for Physicists: {{A}} Tutorial Essay on Control},
  author = {Bechhoefer, John},
  year = {2005},
  journal = {Reviews of Modern Physics},
  volume = {77},
  number = {3},
  pages = {783--836},
  issn = {00346861},
  doi = {10.1103/RevModPhys.77.783}
}

@article{Biance2009,
  title = {Topological Transition Dynamics in a Strained Bubble Cluster},
  author = {Biance, Anne-Laure and {Cohen-Addad}, Sylvie and H{\"o}hler, Reinhard},
  year = {2009},
  journal = {Soft Matter},
  volume = {5},
  number = {23},
  pages = {4672},
  issn = {1744-683X, 1744-6848},
  doi = {10.1039/b910150k},
  urldate = {2024-07-19},
  langid = {english}
}

@article{Bitzek2006,
  title = {Structural Relaxation Made Simple},
  author = {Bitzek, Erik and Koskinen, Pekka and G{\"a}hler, Franz and Moseler, Michael and Gumbsch, Peter},
  year = {2006},
  journal = {Physical Review Letters},
  volume = {97},
  number = {17},
  pages = {1--4},
  issn = {00319007},
  doi = {10.1103/PhysRevLett.97.170201},
  isbn = {0031-9007},
  pmid = {17155444}
}

@article{Boattini2020,
  title = {Autonomously Revealing Hidden Local Structures in Supercooled Liquids},
  author = {Boattini, Emanuele and {Mar{\'i}n-Aguilar}, Susana and Mitra, Saheli and Foffi, Giuseppe and Smallenburg, Frank and Filion, Laura},
  year = {2020},
  month = oct,
  journal = {Nature Communications},
  volume = {11},
  number = {1},
  pages = {5479},
  publisher = {Nature Publishing Group},
  issn = {2041-1723},
  doi = {10.1038/s41467-020-19286-8},
  urldate = {2024-07-23},
  copyright = {2020 The Author(s)},
  langid = {english}
}

@article{Brogliato2003,
  title = {Some Perspectives on the Analysis and Control of Complementarity Systems},
  author = {Brogliato, Bernard},
  year = {2003},
  journal = {IEEE Transactions on Automatic Control},
  volume = {48},
  number = {6},
  pages = {918--935},
  doi = {10.1109/TAC.2003.812777}
}

@article{Cai2021,
  title = {Mesoscale Metrics on Approach to the Clogging Point},
  author = {Cai, Grace and Harada, Anna Belle and Nordstrom, Kerstin},
  year = {2021},
  month = aug,
  journal = {Granular Matter},
  volume = {23},
  number = {3},
  pages = {69},
  issn = {1434-5021, 1434-7636},
  doi = {10.1007/s10035-021-01133-2},
  urldate = {2024-07-19},
  langid = {english}
}

@article{Cates1998,
  title = {Jamming, {{Force Chains}}, and {{Fragile Matter}}},
  author = {Cates, Michael E. and Wittmer, J. P. and Bouchaud, J.-P. and Claudin, Philippe},
  year = {1998},
  month = aug,
  journal = {Physical Review Letters},
  volume = {81},
  number = {9},
  pages = {1841--1844},
  publisher = {American Physical Society},
  doi = {10.1103/PhysRevLett.81.1841},
  urldate = {2024-07-18}
}

@article{Chen2010c,
  title = {Microscopic Structural Relaxation in a Sheared Supercooled Colloidal Liquid},
  author = {Chen, Dandan and Semwogerere, Denis and Sato, Jun and Breedveld, Victor and Weeks, Eric R.},
  year = {2010},
  journal = {Physical Review E - Statistical, Nonlinear, and Soft Matter Physics},
  volume = {81},
  number = {1},
  eprint = {0908.4226},
  pages = {1--13},
  issn = {15393755},
  doi = {10.1103/PhysRevE.81.011403},
  archiveprefix = {arXiv}
}

@article{Chen2011,
  title = {Measurement of Correlations between Low-Frequency Vibrational Modes and Particle Rearrangements in Quasi-Two-Dimensional Colloidal Glasses},
  author = {Chen, Ke and Manning, M. L. and Yunker, Peter J. and Ellenbroek, Wouter G. and Zhang, Zexin and Liu, Andrea J. and Yodh, Arjun G.},
  year = {2011},
  journal = {Physical Review Letters},
  volume = {107},
  number = {10},
  eprint = {1103.2352},
  pages = {1--5},
  issn = {00319007},
  doi = {10.1103/PhysRevLett.107.108301},
  archiveprefix = {arXiv}
}

@article{Chen2018,
  title = {Stress Anisotropy in Shear-Jammed Packings of Frictionless Disks},
  author = {Chen, Sheng and Bertrand, Thibault and Jin, Weiwei and Shattuck, Mark D. and O'Hern, Corey S.},
  year = {2018},
  month = oct,
  journal = {Physical Review E},
  volume = {98},
  number = {4},
  eprint = {1804.10962},
  pages = {042906},
  publisher = {American Physical Society},
  issn = {2470-0045},
  doi = {10.1103/PhysRevE.98.042906},
  archiveprefix = {arXiv}
}

@article{Cubuk2015,
  title = {Identifying {{Structural Flow Defects}} in {{Disordered Solids Using Machine-Learning Methods}}},
  author = {Cubuk, Ekin D. and Schoenholz, S. S. and Rieser, J. M. and Malone, B. D. and Rottler, J. and Durian, D. J. and Kaxiras, E. and Liu, Andrea J.},
  year = {2015},
  month = mar,
  journal = {Physical Review Letters},
  volume = {114},
  number = {10},
  pages = {108001},
  publisher = {American Physical Society},
  doi = {10.1103/PhysRevLett.114.108001},
  urldate = {2024-07-18}
}

@article{Dagois-Bohy2012,
  title = {Soft-Sphere Packings at Finite Pressure but Unstable to Shear},
  author = {{Dagois-Bohy}, Simon and Tighe, Brian P. and Simon, Johannes and Henkes, Silke and Van Hecke, Martin},
  year = {2012},
  journal = {Physical Review Letters},
  volume = {109},
  number = {9},
  eprint = {1203.3364},
  pages = {1--5},
  issn = {00319007},
  doi = {10.1103/PhysRevLett.109.095703},
  archiveprefix = {arXiv},
  pmid = {23002855}
}

@article{Debregeas2001,
  title = {Deformation and {{Flow}} of a {{Two-Dimensional Foam}} under {{Continuous Shear}}},
  author = {Debr{\'e}geas, Georges and Tabuteau, H. and {di Meglio}, Jean-Marc},
  year = {2001},
  month = oct,
  journal = {Physical Review Letters},
  volume = {87},
  number = {17},
  pages = {178305},
  publisher = {American Physical Society},
  doi = {10.1103/PhysRevLett.87.178305},
  urldate = {2024-07-19}
}

@article{Desmond2015,
  title = {Measurement of {{Stress Redistribution}} in {{Flowing Emulsions}}},
  author = {Desmond, Kenneth W. and Weeks, Eric R.},
  year = {2015},
  month = aug,
  journal = {Physical Review Letters},
  volume = {115},
  number = {9},
  pages = {098302},
  publisher = {American Physical Society},
  doi = {10.1103/PhysRevLett.115.098302},
  urldate = {2024-07-18}
}

@article{Dong2023,
  title = {Non-Affine Atomic Rearrangement of Glasses through Stress-Induced Structural Anisotropy},
  author = {Dong, Jie and Peng, Hailong and Wang, Hui and Tong, Yang and Wang, Yutian and Dmowski, Wojciech and Egami, Takeshi and Sun, Baoan and Wang, Weihua and Bai, Haiyang},
  year = {2023},
  month = dec,
  journal = {Nature Physics},
  volume = {19},
  number = {12},
  pages = {1896--1903},
  issn = {1745-2473, 1745-2481},
  doi = {10.1038/s41567-023-02243-9},
  urldate = {2024-07-18},
  langid = {english}
}

@article{Falk1998,
  title = {Dynamics of Viscoplastic Deformation in Amorphous Solids},
  author = {Falk, Michael L and Langer, James S},
  year = {1998},
  month = jun,
  journal = {Physical Review E},
  volume = {57},
  number = {6},
  pages = {7192--7205},
  issn = {1063-651X},
  doi = {10.1103/PhysRevE.57.7192}
}

@article{Fan2017a,
  title = {Particle Rearrangement and Softening Contributions to the Nonlinear Mechanical Response of Glasses},
  author = {Fan, Meng and Zhang, Kai and Schroers, Jan and Shattuck, Mark D. and O'Hern, Corey S.},
  year = {2017},
  journal = {Physical Review E},
  volume = {96},
  number = {3},
  eprint = {1705.06374},
  pages = {1--18},
  issn = {24700053},
  doi = {10.1103/PhysRevE.96.032602},
  archiveprefix = {arXiv},
  pmid = {29346996}
}

@article{Ferdowsi2017,
  title = {Glassy Dynamics of Landscape Evolution},
  author = {Ferdowsi, Behrooz and Ortiz, Carlos P. and Jerolmack, Douglas J.},
  year = {2018},
  journal = {Proceedings of the National Academy of Sciences},
  volume = {115},
  number = {19},
  eprint = {1708.06032},
  pages = {4827--4832},
  issn = {0027-8424},
  doi = {10.1073/pnas.1715250115},
  archiveprefix = {arXiv},
  isbn = {1-4244-0791-5},
  pmid = {21460107}
}

@article{Flechsig2017,
  title = {Design of Elastic Networks with Evolutionary Optimized Long-Range Communication as Mechanical Models of Allosteric Proteins},
  author = {Flechsig, Holger},
  year = {2017},
  journal = {Biophysical Journal},
  volume = {113},
  number = {3},
  eprint = {1702.08317},
  pages = {558--571},
  publisher = {Biophysical Society},
  issn = {15420086},
  doi = {10.1016/j.bpj.2017.06.043},
  archiveprefix = {arXiv},
  pmid = {28793211}
}

@article{Flechsig2018,
  title = {Designed Elastic Networks: {{Models}} of Complex Protein Machinery},
  author = {Flechsig, Holger and Togashi, Yuichi},
  year = {2018},
  journal = {International Journal of Molecular Sciences},
  volume = {19},
  number = {10},
  issn = {14220067},
  doi = {10.3390/ijms19103152},
  pmid = {30322149}
}

@article{Gao2006,
  title = {Frequency Distribution of Mechanically Stable Disk Packings},
  author = {Gao, Guo Jie and B{\l}awzdziewicz, Jerzy and O'Hern, Corey S.},
  year = {2006},
  journal = {Physical Review E - Statistical, Nonlinear, and Soft Matter Physics},
  volume = {74},
  number = {6},
  pages = {1--20},
  issn = {15393755},
  doi = {10.1103/PhysRevE.74.061304}
}

@article{Gartner2016,
  title = {Nonlinear Plastic Modes in Disordered Solids},
  author = {Gartner, Luka and Lerner, Edan},
  year = {2016},
  month = jan,
  journal = {Physical Review E},
  volume = {93},
  number = {1},
  pages = {011001},
  publisher = {American Physical Society},
  doi = {10.1103/PhysRevE.93.011001},
  urldate = {2024-07-23}
}

@article{Goodrich2014b,
  title = {Jamming in Finite Systems: {{Stability}}, Anisotropy, Fluctuations, and Scaling},
  author = {Goodrich, Carl P. and {Dagois-Bohy}, Simon and Tighe, Brian P. and Van Hecke, Martin and Liu, Andrea J. and Nagel, Sidney R.},
  year = {2014},
  journal = {Physical Review E - Statistical, Nonlinear, and Soft Matter Physics},
  volume = {90},
  number = {2},
  eprint = {1406.1529},
  pages = {1--17},
  issn = {15502376},
  doi = {10.1103/PhysRevE.90.022138},
  archiveprefix = {arXiv},
  pmid = {25215719}
}

@article{Gottheil2023,
  title = {State of {{Cell Unjamming Correlates}} with {{Distant Metastasis}} in {{Cancer Patients}}},
  author = {Gottheil, Pablo and Lippoldt, J{\"u}rgen and Grosser, Steffen and Renner, Fr{\'e}d{\'e}ric and Saibah, Mohamad and Tschodu, Dimitrij and Po{\ss}{\"o}gel, Anne-Kathrin and Wegscheider, Anne-Sophie and Ulm, Bernhard and Friedrichs, Kay and Lindner, Christoph and Engel, Christoph and L{\"o}ffler, Markus and Wolf, Benjamin and H{\"o}ckel, Michael and Aktas, Bahriye and Kubitschke, Hans and Niendorf, Axel and K{\"a}s, Josef A.},
  year = {2023},
  month = jul,
  journal = {Physical Review X},
  volume = {13},
  number = {3},
  pages = {031003},
  publisher = {American Physical Society},
  doi = {10.1103/PhysRevX.13.031003},
  urldate = {2024-07-22}
}

@article{Gu2015,
  title = {Controllability of Structural Brain Networks},
  author = {Gu, Shi and Pasqualetti, Fabio and Cieslak, Matthew and Telesford, Qawi K. and Yu, Alfred B. and Kahn, Ari E. and Medaglia, John D. and Vettel, Jean M. and Miller, Michael B. and Grafton, Scott T. and Bassett, Danielle S.},
  year = {2015},
  journal = {Nature Communications},
  volume = {6},
  eprint = {1406.5197},
  pages = {1--10},
  publisher = {Nature Publishing Group},
  issn = {20411723},
  doi = {10.1038/ncomms9414},
  archiveprefix = {arXiv},
  isbn = {2041-1723 (Electronic){\textbackslash}r2041-1723 (Linking)},
  pmid = {26423222}
}

@article{Hayman2011,
  title = {Granular {{Controls}} on {{Periodicity}} of {{Stick-Slip Events}}: {{Kinematics}} and {{Force-Chains}} in an {{Experimental Fault}}},
  shorttitle = {Granular {{Controls}} on {{Periodicity}} of {{Stick-Slip Events}}},
  author = {Hayman, Nicholas W. and Duclou{\'e}, Lucie and Foco, Kate L. and Daniels, Karen E.},
  year = {2011},
  month = dec,
  journal = {Pure and Applied Geophysics},
  volume = {168},
  number = {12},
  pages = {2239--2257},
  issn = {0033-4553, 1420-9136},
  doi = {10.1007/s00024-011-0269-3},
  urldate = {2024-07-19},
  copyright = {http://www.springer.com/tdm},
  langid = {english}
}

@article{Hogan2022,
  title = {Contact and {{Physical Interaction}}},
  author = {Hogan, Neville},
  year = {2022},
  month = may,
  journal = {Annual Review of Control, Robotics, and Autonomous Systems},
  volume = {5},
  number = {1},
  pages = {179--203},
  issn = {2573-5144, 2573-5144},
  doi = {10.1146/annurev-control-042920-010933},
  urldate = {2024-07-15},
  langid = {english}
}

@book{Iglesias2010,
  title = {Control {{Theory}} and {{Systems Biology}}},
  author = {Iglesias, Pablo A. and Ingalls, Brian P.},
  year = {2010},
  publisher = {MIT Press},
  googlebooks = {dojJ6QrTugoC},
  isbn = {978-0-262-01334-5},
  langid = {english}
}

@article{Jensen2014,
  title = {Local Shear Transformations in Deformed and Quiescent Hard-Sphere Colloidal Glasses},
  author = {Jensen, Katharine E. and Weitz, D. A. and Spaepen, Frans},
  year = {2014},
  month = oct,
  journal = {Physical Review E},
  volume = {90},
  number = {4},
  pages = {042305},
  publisher = {American Physical Society},
  doi = {10.1103/PhysRevE.90.042305},
  urldate = {2024-07-18}
}

@article{Jerolmack2019,
  title = {Viewing {{Earth}}'s Surface as a Soft-Matter Landscape},
  author = {Jerolmack, Douglas J and Daniels, Karen E},
  year = {2019},
  journal = {Nature Reviews Physics},
  volume = {1},
  number = {12},
  pages = {716--730},
  issn = {2522-5820},
  doi = {10.1038/s42254-019-0111-x}
}

@book{Kailath1980,
  title = {Linear Systems},
  author = {Kailath, Thomas},
  year = {1980},
  series = {Prentice-{{Hall}} Information and System Sciences Series},
  publisher = {Prentice-Hall},
  address = {Englewood Cliffs, NJ},
  isbn = {978-0-13-536961-6},
  langid = {english}
}

@article{Kapteijns2020,
  title = {Nonlinear Quasilocalized Excitations in Glasses: {{True}} Representatives of Soft Spots},
  shorttitle = {Nonlinear Quasilocalized Excitations in Glasses},
  author = {Kapteijns, Geert and Richard, David and Lerner, Edan},
  year = {2020},
  month = mar,
  journal = {Physical Review E},
  volume = {101},
  number = {3},
  pages = {032130},
  publisher = {American Physical Society},
  doi = {10.1103/PhysRevE.101.032130},
  urldate = {2024-07-23}
}

@article{Katgert2013,
  title = {The Jamming Perspective on Wet Foams},
  author = {Katgert, Gijs and Tighe, Brian P. and Van Hecke, Martin},
  year = {2013},
  journal = {Soft Matter},
  volume = {9},
  number = {41},
  pages = {9739},
  issn = {1744-683X, 1744-6848},
  doi = {10.1039/c3sm51543e},
  urldate = {2024-07-18},
  langid = {english}
}

@article{Kawasaki2007,
  title = {Correlation between Dynamic Heterogeneity and Medium-Range Order in Two-Dimensional Glass-Forming Liquids},
  author = {Kawasaki, Takeshi and Araki, Takeaki and Tanaka, Hajime},
  year = {2007},
  journal = {Physical Review Letters},
  volume = {99},
  number = {21},
  pages = {215701},
  issn = {00319007},
  doi = {10.1103/PhysRevLett.99.215701},
  isbn = {0031-9007},
  pmid = {18233228}
}

@article{Keim2013,
  title = {Yielding and Microstructure in a {{2D}} Jammed Material under Shear Deformation},
  author = {Keim, Nathan C. and Arratia, Paulo E.},
  year = {2013},
  journal = {Soft Matter},
  volume = {9},
  number = {27},
  eprint = {1304.2253},
  pages = {6222--6225},
  issn = {1744683X},
  doi = {10.1039/c3sm51014j},
  archiveprefix = {arXiv},
  isbn = {1744-683X},
  pmid = {24173598}
}

@article{Keim2014,
  title = {Mechanical and {{Microscopic Properties}} of the {{Reversible Plastic Regime}} in a {{2D Jammed Material}}},
  author = {Keim, Nathan C. and Arratia, Paulo E.},
  year = {2014},
  journal = {Physical Review Letters},
  volume = {112},
  number = {2},
  eprint = {1308.6806},
  pages = {1--5},
  issn = {00319007},
  doi = {10.1103/PhysRevLett.112.028302},
  archiveprefix = {arXiv},
  isbn = {1079-7114 (Electronic){\textbackslash}r0031-9007 (Linking)},
  pmid = {24484046}
}

@article{Kim2019,
  title = {Conformational Control of Mechanical Networks},
  author = {Kim, Jason Z. and Lu, Zhixin and Strogatz, Steven H. and Bassett, Danielle S.},
  year = {2019},
  journal = {Nature Physics},
  volume = {15},
  number = {7},
  eprint = {1804.00173},
  pages = {714--720},
  publisher = {Springer US},
  issn = {1745-2473},
  doi = {10.1038/s41567-019-0475-y},
  archiveprefix = {arXiv}
}

@article{Kozlowski2021,
  title = {Stress Propagation in Locally Loaded Packings of Disks and Pentagons},
  author = {Kozlowski, Ryan and Zheng, Hu and Daniels, Karen E. and Socolar, Joshua E. S.},
  year = 2021,
  month = nov,
  journal = {Soft Matter},
  volume = {17},
  number = {44},
  pages = {10120--10127},
  publisher = {The Royal Society of Chemistry},
  issn = {1744-6848},
  doi = {10.1039/D1SM01137E},
  urldate = {2024-07-19},
  langid = {english}
}

@article{Liu2010b,
  title = {The {{Jamming Transition}} and the {{Marginally Jammed Solid}}},
  author = {Liu, Andrea J. and Nagel, Sidney R.},
  year = {2010},
  journal = {Annual Review of Condensed Matter Physics},
  volume = {1},
  number = {1},
  pages = {347--369},
  issn = {1947-5454},
  doi = {10.1146/annurev-conmatphys-070909-104045}
}

@article{Liu2011,
  title = {Controllability of Complex Networks},
  author = {Liu, Yang Yu and Slotine, Jean Jacques and Barab{\'a}si, Albert L{\'a}szl{\'o}},
  year = {2011},
  journal = {Nature},
  volume = {473},
  number = {7346},
  pages = {167--173},
  issn = {00280836},
  doi = {10.1038/nature10011},
  arxiv = {http://www.nature.com/nature/journal/v473/n7346/abs/10.1038-nature10011-unlocked.html\#supplementary-information},
  isbn = {1476-4687 (Electronic){\textbackslash}r0028-0836 (Linking)},
  pmid = {21562557}
}

@article{Malandro1999,
  title = {Relationships of Shear-Induced Changes in the Potential Energy Landscape to the Mechanical Properties of Ductile Glasses},
  author = {Malandro, Dennis L. and Lacks, Daniel J.},
  year = {1999},
  month = mar,
  journal = {The Journal of Chemical Physics},
  volume = {110},
  number = {9},
  pages = {4593--4601},
  issn = {0021-9606},
  doi = {10.1063/1.478340}
}

@article{Malins2013a,
  title = {Identification of Long-Lived Clusters and Their Link to Slow Dynamics in a Model Glass Former},
  author = {Malins, Alex and Eggers, Jens and Royall, C. Patrick and Williams, Stephen R. and Tanaka, Hajime},
  year = {2013},
  journal = {Journal of Chemical Physics},
  volume = {138},
  number = {12},
  issn = {00219606},
  doi = {10.1063/1.4790515},
  isbn = {1089-7690 (Electronic){\textbackslash}r0021-9606 (Linking)},
  pmid = {23556786}
}

@article{Maloney2006,
  title = {Amorphous Systems in Athermal, Quasistatic Shear},
  author = {Maloney, Craig E. and Lemaitre, Anael},
  year = {2006},
  journal = {Physical Review E - Statistical, Nonlinear, and Soft Matter Physics},
  volume = {74},
  number = {1},
  eprint = {cond-mat/0510677},
  pages = {1--22},
  issn = {15393755},
  doi = {10.1103/PhysRevE.74.016118},
  archiveprefix = {arXiv},
  isbn = {1539-3755 (Print){\textbackslash}r1539-3755 (Linking)},
  pmid = {16907162}
}

@article{Manning2011,
  title = {Vibrational {{Modes Identify Soft Spots}} in a {{Sheared Disordered Packing}}},
  author = {Manning, Mary Lisa and Liu, Andrea J.},
  year = {2011},
  month = aug,
  journal = {Physical Review Letters},
  volume = {107},
  number = {10},
  eprint = {1012.4822},
  pages = {108302},
  issn = {0031-9007},
  doi = {10.1103/PhysRevLett.107.108302},
  archiveprefix = {arXiv}
}

@article{Nicolas2018,
  title = {Deformation and Flow of Amorphous Solids: {{Insights}} from Elastoplastic Models},
  author = {Nicolas, Alexandre and Ferrero, Ezequiel E. and Martens, Kirsten and Barrat, Jean Louis},
  year = {2018},
  journal = {Reviews of Modern Physics},
  volume = {90},
  number = {4},
  eprint = {1708.09194},
  pages = {45006},
  publisher = {American Physical Society},
  issn = {15390756},
  doi = {10.1103/RevModPhys.90.045006},
  archiveprefix = {arXiv}
}

@article{Oswald2017,
  title = {Jamming Transitions in Cancer},
  author = {Oswald, Linda and Grosser, Steffen and Smith, David M. and K{\"a}s, Josef A.},
  year = {2017},
  journal = {Journal of Physics D: Applied Physics},
  volume = {50},
  number = {48},
  issn = {13616463},
  doi = {10.1088/1361-6463/aa8e83}
}

@article{Pasqualetti2014,
  title = {Controllability Metrics, Limitations and Algorithms for Complex Networks},
  author = {Pasqualetti, Fabio and Zampieri, Sandro and Bullo, Francesco},
  year = {2014},
  journal = {IEEE Transactions on Control of Network Systems},
  volume = {1},
  number = {1},
  eprint = {1308.1201},
  pages = {40--52},
  issn = {23255870},
  doi = {10.1109/TCNS.2014.2310254},
  archiveprefix = {arXiv},
  isbn = {9781479932726}
}

@article{Patinet2016,
  title = {Connecting {{Local Yield Stresses}} with {{Plastic Activity}} in {{Amorphous Solids}}},
  author = {Patinet, Sylvain and Vandembroucq, Damien and Falk, Michael L.},
  year = {2016},
  journal = {Physical Review Letters},
  volume = {117},
  number = {4},
  pages = {1--5},
  issn = {10797114},
  doi = {10.1103/PhysRevLett.117.045501}
}

@article{Richard2020a,
  title = {Predicting Plasticity in Disordered Solids from Structural Indicators},
  author = {Richard, David and Ozawa, M. and Patinet, S. and Stanifer, E. and Shang, B. and Ridout, S. A. and Xu, B. and Zhang, G. and Morse, P. K. and Barrat, J. L. and Berthier, L. and Falk, M. L. and Guan, P. and Liu, A. J. and Martens, K. and Sastry, S. and Vandembroucq, D. and Lerner, E. and Manning, M. Lisa},
  year = {2020},
  journal = {Physical Review Materials},
  volume = {4},
  number = {11},
  eprint = {2003.11629},
  pages = {1--19},
  issn = {24759953},
  doi = {10.1103/PhysRevMaterials.4.113609},
  archiveprefix = {arXiv}
}

@article{Rocks2017,
  title = {Designing Allostery-Inspired Response in Mechanical Networks},
  author = {Rocks, Jason W. and Pashine, Nidhi and Bischofberger, Irmgard and Goodrich, Carl P. and Liu, Andrea J. and Nagel, Sidney R.},
  year = {2017},
  journal = {Proceedings of the National Academy of Sciences of the United States of America},
  volume = {114},
  number = {10},
  eprint = {1607.08562},
  pages = {2520--2525},
  issn = {10916490},
  doi = {10.1073/pnas.1612139114},
  archiveprefix = {arXiv},
  pmid = {28223534}
}

@article{Schall2007,
  title = {Structural {{Rearrangements That Govern Flow}} in {{Colloidal Glasses}}},
  author = {Schall, Peter and Weitz, David A. and Spaepen, Frans},
  year = {2007},
  month = dec,
  journal = {Science},
  volume = {318},
  number = {5858},
  pages = {1895--1899},
  publisher = {American Association for the Advancement of Science},
  doi = {10.1126/science.1149308},
  urldate = {2024-07-18}
}

@article{Schoenholz2017,
  title = {Combining {{Machine Learning}} and {{Physics}} to {{Understand Glassy Systems}}},
  author = {Schoenholz, Samuel S.},
  year = 2018,
  month = jun,
  journal = {Journal of Physics: Conference Series},
  volume = {1036},
  number = {1},
  pages = {012021},
  publisher = {IOP Publishing},
  issn = {1742-6596},
  doi = {10.1088/1742-6596/1036/1/012021},
  urldate = {2026-03-06},
  langid = {english}
}

@book{Schulz2006,
  title = {Control Theory in Physics and Other Fields of Science: Concepts, Tools, and Applications},
  shorttitle = {Control Theory in Physics and Other Fields of Science},
  author = {Schulz, Michael},
  year = {2006},
  series = {Springer Tracts in Modern Physics},
  number = {v. 215},
  publisher = {Springer},
  address = {Berlin ; New York},
  isbn = {978-3-540-29514-3},
  langid = {english},
  lccn = {QA402.3 .S355 2006}
}

@article{Shang2014,
  title = {Evolution of Atomic Rearrangements in Deformation in Metallic Glasses},
  author = {Shang, Baoshuang S. and Li, M. Z. and Yao, Y. G. and Lu, Y. J. and Wang, Weihua H.},
  year = {2014},
  month = oct,
  journal = {Physical Review E},
  volume = {90},
  number = {4},
  pages = {042303},
  publisher = {American Physical Society},
  doi = {10.1103/PhysRevE.90.042303},
  urldate = {2024-07-18}
}

@article{Spaepen1977,
  title = {A Microscopic Mechanism for Steady State Inhomogeneous Flow in Metallic Glasses},
  author = {Spaepen, Frans},
  year = {1977},
  month = apr,
  journal = {Acta Metallurgica},
  volume = {25},
  number = {4},
  pages = {407--415},
  issn = {00016160},
  doi = {10.1016/0001-6160(77)90232-2},
  urldate = {2024-07-18},
  copyright = {https://www.elsevier.com/tdm/userlicense/1.0/},
  langid = {english}
}

@article{Summers2014,
  title = {Optimal {{Sensor}} and {{Actuator Placement}} in {{Complex Dynamical Networks}}},
  author = {Summers, Tyler H. and Lygeros, John},
  year = {2014},
  month = jan,
  journal = {IFAC Proceedings Volumes},
  series = {19th {{IFAC World Congress}}},
  volume = {47},
  number = {3},
  pages = {3784--3789},
  issn = {1474-6670},
  doi = {10.3182/20140824-6-ZA-1003.00226},
  urldate = {2024-07-24}
}

@article{Tanguy2006,
  title = {Plastic Response of a {{2D Lennard-Jones}} Amorphous Solid: {{Detailed}} Analysis of the Local Rearrangements at Very Slow Strain Rate},
  author = {Tanguy, Anne and Leonforte, F. and Barrat, Jean-Louis},
  year = {2006},
  journal = {European Physical Journal E},
  volume = {20},
  number = {3},
  eprint = {cond-mat/0605397},
  pages = {355--364},
  issn = {12928941},
  doi = {10.1140/epje/i2006-10024-2},
  archiveprefix = {arXiv}
}

@article{Tanguy2010,
  title = {Vibrational Modes as a Predictor for Plasticity in a Model Glass},
  author = {Tanguy, Anne and Mantisi, B. and Tsamados, Michel},
  year = {2010},
  month = apr,
  journal = {EPL (Europhysics Letters)},
  volume = {90},
  number = {1},
  pages = {16004},
  issn = {0295-5075, 1286-4854},
  doi = {10.1209/0295-5075/90/16004},
  urldate = {2024-07-19},
  langid = {english}
}

@article{Tong2014,
  title = {Order Parameter for Structural Heterogeneity in Disordered Solids},
  author = {Tong, Hua and Xu, Ning},
  year = {2014},
  journal = {Physical Review E - Statistical, Nonlinear, and Soft Matter Physics},
  volume = {90},
  number = {1},
  pages = {1--5},
  issn = {15502376},
  doi = {10.1103/PhysRevE.90.010401},
  pmid = {25122238}
}

@article{Tong2018,
  title = {Revealing {{Hidden Structural Order Controlling Both Fast}} and {{Slow Glassy Dynamics}} in {{Supercooled Liquids}}},
  author = {Tong, Hua and Tanaka, Hajime},
  year = {2018},
  month = mar,
  journal = {Physical Review X},
  volume = {8},
  number = {1},
  pages = {011041},
  publisher = {American Physical Society},
  issn = {2160-3308},
  doi = {10.1103/PhysRevX.8.011041}
}

@article{Widmer-Cooper2008,
  title = {Irreversible Reorganization in a Supercooled Liquid Originates from Localized Soft Modes},
  author = {{Widmer-Cooper}, Asaph and Perry, Heidi and Harrowell, Peter and Reichman, David R.},
  year = {2008},
  month = sep,
  journal = {Nature Physics},
  volume = {4},
  number = {9},
  pages = {711--715},
  publisher = {Nature Publishing Group},
  issn = {1745-2481},
  doi = {10.1038/nphys1025},
  urldate = {2024-07-18},
  copyright = {2008 Springer Nature Limited},
  langid = {english}
}

@book{Wit2012,
  title = {Theory of {{Robot Control}}},
  author = {de Wit, Carlos Canudas and Siciliano, Bruno and Bastin, Georges},
  year = {2012},
  month = dec,
  publisher = {Springer Science \& Business Media},
  googlebooks = {5NnUBwAAQBAJ},
  isbn = {978-1-4471-1501-4},
  langid = {english}
}

\end{document}